\documentclass[aps,preprint,amssymb,superscriptaddress]{revtex4-1}

\usepackage{floatrow}
\newfloatcommand{capbtabbox}{table}[\FBwidth]
\usepackage{graphicx}% Include figure files
\usepackage{dcolumn}% Align table columns on decimal point
\usepackage{bm}% bold math
\usepackage{natbib}
\usepackage{CJK}
\usepackage[utf8]{inputenc}
\usepackage[T1]{fontenc}
\usepackage{mathptmx}
\usepackage{gensymb}
\usepackage{graphicx}% Include figure files
\usepackage{epstopdf}
\usepackage{dcolumn}% Align table columns on decimal point
\usepackage{bm}% bold math
\usepackage{amsmath}
\usepackage{mathtools}
\usepackage{relsize}
\usepackage{multirow}
\usepackage[colorlinks]{hyperref}

\begin{document}

%\preprint{AIP/123-QED}

%\begin{CJK*}{GB}{}
%\preprint{APS/123-QED}
\title{Theoretical Investigation of Yield-Enhancing Equilibrium Negatively Ionized Tin-Vacancy Center Preparation Pathways in N-Doped Diamond}% Force line breaks with \\
\author{Aditya Bahulikar}
\affiliation{Department of Electrical Engineering and Computer Science, Syracuse University, Syracuse, NY 13210}
\author{Steven L. Richardson} 
\affiliation{John A. Paulson School of Engineering and Applied Sciences, Harvard University, Cambridge, MA 02138, USA}
\affiliation{Department of Electrical and Computer Engineering, Howard University, Washington, DC 20059}
\author{Rodrick Kuate Defo}
\email{rkuatede@syr.edu}
\affiliation{Department of Electrical Engineering and Computer Science, Syracuse University, Syracuse, NY 13210}
\date{\today}% It is always \today, today,

\begin{abstract}

The elucidation of the mechanism of Sn$V^-$ formation in diamond is especially important as the Sn$V^-$ color center has the potential to be a superior single-photon emitter when compared to the N$V$ and to other Group IV color centers. The typical formation of the Sn$V$ involves placing Sn in diamond by ion implantation, but the formation of a charged Sn$V$ species requires an additional complication. This complication is related to the energy cost associated with electronic transitions within the host diamond. Effectively, producing the Sn$V^-$ charge state using an electron obtained from a band edge of the host diamond is less energetically favorable than having the Sn$V^-$ receive an electron from a neighboring donor dopant. Among donor dopants, substitutional N (N$_\text{C}$) is always present in even the purest synthetic or natural diamond sample. The mechanism of electron donation by N$_\text{C}$ has been proposed by Collins for charging the N$V$ in diamond and it has been used to interpret many experimental results. Therefore, in this paper we use DFT to explore the pathways for the formation of the Sn$V^-$ charge state due to electron donation arising from the presence of N$_\text{C}$ in the host diamond. Explicitly, defect concentrations are calculated in equilibrium in each of the explored pathways to determine the yield of the Sn$V^-$ throughout each of the pathways. The importance of our work is to suggest experimental ways of enhancing the yield of charged states like the Sn$V^-$ in diamond for transformative applications in optoelectronics and quantum information.

\end{abstract}

\maketitle

\section{INTRODUCTION}

The structure of the charged tin-vacancy color center Sn$V^-$ in diamond was first proposed by Goss {\it et al.}~\cite{goss1996twelve,goss2007density,goss2005vacancy} through first-principles density-functional theory calculations using the local density approximation (LDA). This color center was first experimentally realized via ion implantation followed by high-temperature annealing~\cite{Iwasaki2, tchernij2017single}, an approach also used for Pb$V^-$~\cite{Trusheim,wang_low_2021} and Si$V^-$~\cite{lai_high_2022}. It was subsequently generated by microwave plasma chemical vapor deposition~\cite{westerhausen2020controlled} and has emerged as a very important candidate for a solid state single photon emitter. The direct identification and quantification of the atomic configurations of Sn-related centers in diamond, including those that are not optically active, has been determined by $\beta$ emission channeling following low fluence $^{121}$Sn implantation~\cite{wahl2020direct}. Unlike the N$V^-$ and Si$V^-$ color centers, the Sn$V^-$ color center has the advantage of both a long spin coherence time above millikelvin temperatures ~\cite{wahl2020direct,kuruma2021coupling,rosenthal2024single,rosenthal2023microwave} and a large Debye-Waller factor~\cite{wahl2020direct,Trusheim2020}. Subsequent work has demonstrated the utility of the Sn$V^-$ in diamond in a variety of applications including luminescent thermometry~\cite{Alkahtani2018tin}, high-temperature high-pressure studies~\cite{Ekimov2018tin,palyanov2019high,fukuta2021sn}, which is also applicable to Ge$V^-$ centers~\cite{Palyanov}, and potentially as an optically accessible quantum memory~\cite{rugar2019char}. The optical and spin properties of Sn$V^-$ color centers in diamond have also been explored through cavity enhancement~\cite{Rugar2021, kuruma2021coupling}, cryogenic magneto-optical and spin spectroscopy~\cite{Trusheim2020}, photoluminescence excitation spectroscopy~\cite{Gorlitz2020}, all-optical Ramsey interferometry~\cite{debroux2021quantum}, coherent population trapping~\cite{gorlitz2022coherence,debroux2021quantum}, charge-state tuning~\cite{luehmann2020charge}, and the Stark effect~\cite{de2021investigation, aghaeimeibodi2021electrical}. Theoretical and computational work on Sn$V^-$ has focused on complexes of Sn$V^-$ with $V_{\text{C}}$~\cite{Kuate8}, the hyperfine tensor of Sn$V^-$~\cite{defo2021calculating}, the Jahn-Teller effect in Sn$V^-$~\cite{ciccarino2020strong}, and the magneto-optical spectrum of Sn$V^-$~\cite{Gali3}. Investigations have also been done of color centers in other host systems such as the polytypes of SiC and silicon~\cite{Kraus2,Castelletto,Koehl,Riedel,Kimoto,Wang17prap,Wang17acsp,Fuchs,Kuate,Gadalla2021enhanced,DEFO2019,Nagy2018quantum,Kraus17,Widmann,Lohrmann,Bracher,Falk,Soykal,Soykal2,Weber,Kraus,Gali,awschalom2018quant,Whiteley2019spin,Wolfowicz2021quantum,kuate_theoretical_2023,kuate_charge_2023,kuate_investigating_2024,day_electrical_2024}. 

In ion implantation experiments, Coulomb-driven defect engineering has achieved the highest yield for the Sn$V^-$ in diamond~\cite{luhmann2019coulomb}, following pioneering work with N$V$ centers in diamond~\cite{Oliveira}. Specifically, $n$-type doping leads to the highest yield for the Sn$V^-$ compared to $p$-type doping or to no doping, in agreement with theoretical results that investigated the formation energy as a function of Fermi level when Sn$V^-$ and a single vacancy were attached, isolated or separated by a finite distance~\cite{Kuate8}. Nonetheless, such yields are still well below those of the Si$V^-$ and N$V^-$~\cite{Bradac2019}. Therefore, we investigate the effect of modulating the chemical potentials for N and Sn as the temperature is lowered in the sample preparation process in order to identify pathways leading both to a high yield and to an appreciable defect concentration for the Sn$V^-$ in N-doped diamond. We model Sn$V$ in the presence of substitutional N (N$_\text{C}$) due to the abundance of N$_\text{C}$ in diamond and to the fact that electron transfer from N$_\text{C}$ to Sn$V$ centers is energetically more favorable than obtaining an electron from the band edges of the host. In Section \ref{sec:methods}, we present the theoretical approach used in this work as well as a discussion of the theoretical results and we conclude in Section \ref{sec:conc}.

\section{THEORETICAL APPROACH AND DISCUSSION \label{sec:methods}}
\subsection{Theory \label{sec:methodintro}}
We begin our elucidation of the theoretical formalism with the equation for the formation energy $\Delta H_f$ of a species, X$^{\rm q}$,~\cite{Vinichenko,zhang1991chemical, RevModPhys.86.253,Kuate8,Kuate2021methods,Kuate,zunger2021under,Yang2015,ashcroft1976solid,kuate_theoretical_2023}
\begin{equation}
\label{eq:form_eq}
\Delta H_f({\rm X^{\rm q}}, \, \{\mu_i^\text{X}\}, \, E_\text{F}) = E_{\text{def}}({\rm X^{\rm q}}) - E_0 - \sum_i\mu_i^\text{X}n_i + {\rm q}\, E_{\text{F}} + E_{\text{corr}}(\rm X^{\rm q}),
\end{equation}
where $E_{\text{def}}({\rm X^{\rm q}})$ is the energy of the charged supercell with the X$^{\rm q}$ species, $E_0$ is the energy of the stoichiometric neutral supercell, $\mu_i^\text{X}$ is the chemical potential of the $i^{\rm th}$ species that was removed or added to produce the supercell with the X$^{\rm q}$ species, $n_i$ is a positive (negative) integer representing the number of the $i^{\rm th}$ species that was added (removed), q represents the charge state of the species X$^{\rm q}$, and $E_{\text{F}}$ is the absolute position of the Fermi level and is treated as a parameter. A correction term to account for a finite supercell when performing a calculation for a charged defect is included in the formation energy equation as well~\cite{Vinichenko,Freysoldt2011electrostatic,Freysoldt2009fully,Kumagai,Komsa2013finite,Walsh2021} and is labeled $E_{\text{corr}}(\rm X^{\rm q})$. 

We note that a defect species has a formation energy which depends on $E_\text{F}$ and which is lowest for any particular charge state of that defect species over a single range of $E_\text{F}$. This fact allows us to define a charge-transition level $\epsilon^{\rm X}(\text{q}/\text{q}^{\prime})$ (CTL) which is defined as the value of the Fermi level $E_{\text{F}}^*$ for which X$^\text{q}$ and $\text{X}^{\text{q}^{\prime}}$ have equivalent formation energies or $\Delta H_f({\rm X^{\rm q}}, \, \mu_i, \, E_\text{F}^*)$  = $\Delta H_f({\rm X^{\rm q^{\prime} }}, \, \{\mu_i^\text{X}\}, \, E_\text{F}^*)$. Using Eq. (\ref{eq:form_eq}) we can explicitly define our CTL  as~\cite{RevModPhys.86.253,kuate_theoretical_2023}
\begin{equation}
\label{eq:CTL_eq} 
\epsilon^{\rm{X}} (\text{q}/\text{q}^{\prime}) = \frac{(E_{\text{def}}({\rm X^{\rm q}}) +E_{\text{corr}}({\rm X^{\rm q}})) - (E_{\text{def}}({\rm X^{\rm q^{\prime}}}) +E_{\text{corr}}({\rm X^{\rm q^{\prime}}}))}{\text{q}^{\prime} - \text{q}}.
\end{equation}
Since the set of charge-transition levels of a defect and the formation energy for a single charge state of the defect are sufficient to determine the formation energies for the remaining charge states of the defect, it suffices to compute the formation energy for the neutral charge state of the defect if the charge-transition levels of the defect of interest are known. 

Given the formation energies for the various charge states of a set of defect species, one can determine their concentrations through the equation~\cite{BUCKERIDGE2019329,RevModPhys.86.253}

\begin{equation}
\label{eq:model-1}
    n_{\rm X^{\rm q}}(\left\{\mu_i^{\text{X}}\right\}, \, E_\text{F}) = N_{\text{X}}\,g_{\text{X}^{\text{q}}}\,\exp(-\Delta H_f({\rm X^{\rm q}}, \, \{\mu_i^\text{X}\}, \, E_\text{F})/k_BT),
\end{equation}
where $T$ denotes temperature, $k_B$ is Boltzmann's constant, $\text{q}$ is the charge state, $N_\text{X}$ is the concentration of crystal sites on which the defect can form, and $g_{\text{X}^{\text{q}}}$ is the degeneracy arising from the symmetry of a given charge state q of the defect.

It is important to realize that in addition to the defect concentration we must consider both the electron concentration $n_0$ and hole concentration $p_0$,  

\begin{equation}
\label{eq:model-2}
    n_0(T) = \int_{E_g}^\infty f_e(E)\rho(E)dE,\\
\end{equation}
and
\begin{equation}
\label{eq:model-3}
    p_0(T) = \int_{-\infty}^0f_h(E)\rho(E)dE.
\end{equation}
where $\rho(E)$ is the density of states per unit volume, $E_g$ is the band gap of the host with the valence band maximum set to zero and all energies shifted accordingly, and the Fermi-Dirac distributions for electrons $f_e$ and holes $f_h$ are given as

\begin{equation}
\label{eq:model-4}
 f_e(E) = \frac{1}{\text{exp}((E-E_{\text{F}})/k_BT)+1}
 \end{equation}
and
\begin{equation}
\label{eq:model-5}
 f_h(E) = 1-f_e(E).
 \end{equation}

So far in this standard approach of computing formation energies for defects, we have assumed that in Eq. (\ref{eq:form_eq}) both $E_\text{F}$ and the set of chemical potentials $
\left\{\mu_i^\text{X}\right\}$ can be treated as independent parameters. Given our extrinsic semiconductor which can be described by an electron carrier concentration $n_0$, a hole carrier concentration $p_0$, and defect concentrations $n_{\text{X}^{\text{q}}}$, we must be assured from physical grounds that the entire system is electrically neutral. For an arbitrary choice of $E_\text{F}$ in  Eq. (\ref{eq:form_eq}) this neutrality condition is not necessarily satisfied. In order to ensure that the neutrality condition is satisfied, for a given choice of the parameter $E_\text{F}$ at a particular value of temperature $T$ and set of chemical potentials $\left\{\mu_i^{\text{X}}\right\}$, we must first evaluate Eq. (\ref{eq:form_eq}) to compute the formation energy  $\Delta H_f({\rm X^{\rm q}}, \, \{\mu_i^\text{X}\}, \, E_\text{F})$. From this formation energy  $\Delta H_f({\rm X^{\rm q}}, \, \{\mu_i^\text{X}\}, \, E_\text{F})$, we can use Eq. (\ref{eq:model-1}) to find the species concentrations $n_{\text{X}^\text{q}}$ and our choice of the parameter $E_\text{F}$ will yield a particular value for $n_0$ and $p_0$. Then, upon imposing  charge neutrality in the crystal it should be true that 

\begin{equation}
    \label{eq:chargebalance}
    n_0 - \sum_{\text{X}}\sum_{\text{q}}\text{q}n_{\text{X}^{\text{q}}} = p_0.
\end{equation}
which will not be initially true. If we repeat this process for a small step from the initial choice of our parameter $E_\text{F}$ we can iterate Eqs. (\ref{eq:form_eq},\ref{eq:model-1},\ref{eq:model-2},\ref{eq:model-3},\ref{eq:chargebalance}) to self-consistency at a given temperature and set of chemical potentials~\cite{BUCKERIDGE2019329,RevModPhys.86.253}. We use the resulting self-consistent or equilibrium $E_{\text{F}}^{eq}$ to calculate the $n_{\text{X}^{\text{q}}}(\left\{\mu_i^{\text{X}}\right\}, T)$, $n_0(\left\{\mu_i^{\text{X}}\right\}, T)$, and $p_0 (\left\{\mu_i^{\text{X}}\right\}, T)$ for a particular temperature $T$ and set of chemical potentials $\left\{\mu_i^{\text{X}}\right\}$. This self-consistent calculation is done using the SC-FERMI code~\cite{BUCKERIDGE2019329} with input from VASP~\cite{Kresse1,Kresse2,Kresse3}.

\subsection{Computational methods \label{sec:methodology}}
 Our density of states and formation energy calculations used VASP~\cite{Kresse1,Kresse2,Kresse3} with the screened hybrid functional of Heyd, Scuseria and Ernzerhof (HSE) employing the original parameters~\cite{Heyd,Krukau}. Our calculations were terminated when the forces in the atomic-position relaxations dropped below a threshold of $10^{-2}$ eV$\cdot$\AA$^{-1}$. The wavefunctions were expanded in a planewave basis with a cutoff energy of 430~eV, the size of the supercell was 512 atoms ($4\times4\times4$ multiple of the conventional unit cell), and gamma-point integration was used. The density of states was computed via the tetrahedron method with Blöchl corrections~\cite{blochl1994imp}. The chemical potentials of all the elements used in our calculations are: N ($\beta$ hexagonal close-packed structure, $-11.39$ eV/atom); Sn (body-centered tetragonal structure, $-4.60$ eV/atom); and C (diamond structure, $-11.28$ eV/atom). 
 
\begin{figure}[ht!] 
\centering
\includegraphics[width=0.5\textwidth]{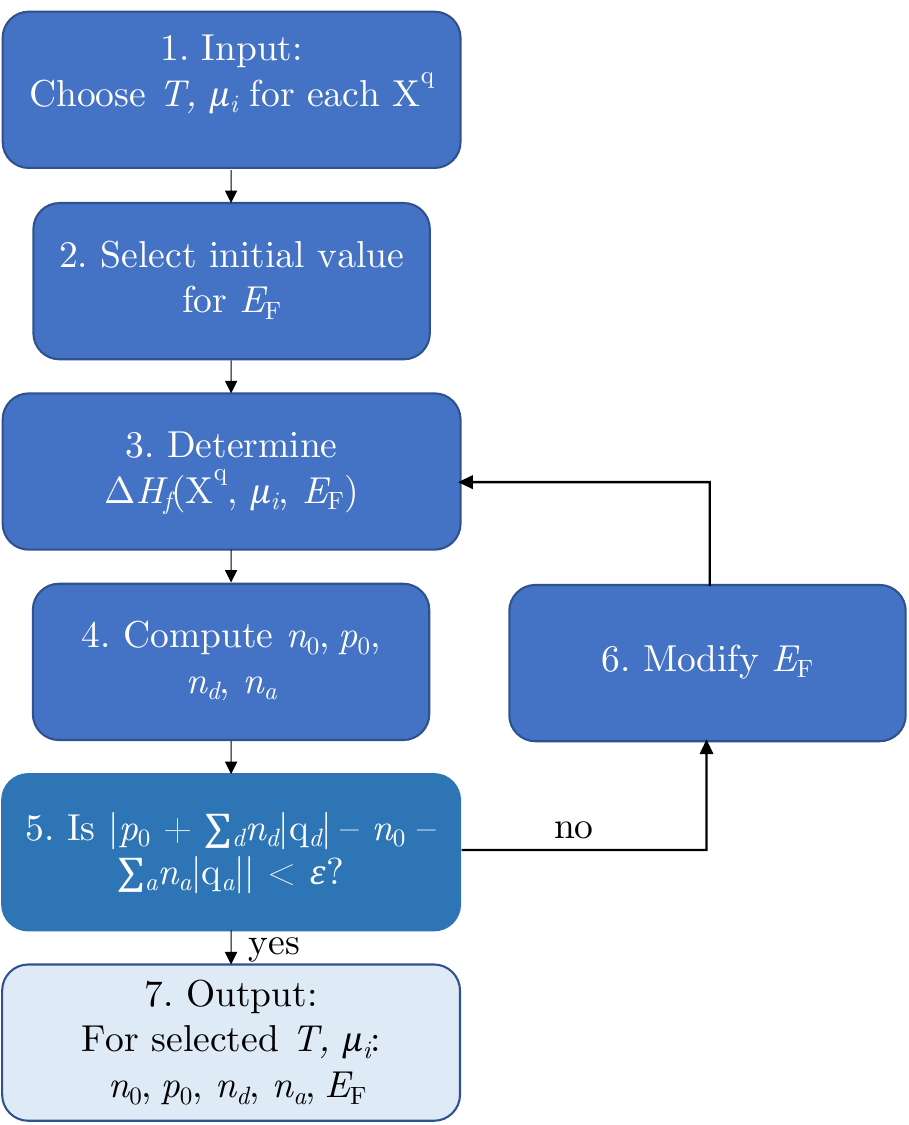}
\caption{Flowchart of the procedure used in this work. The value of $\varepsilon$ was set to $10^{-12}$~eV~\cite{BUCKERIDGE2019329}. The steps 2-7 use the SC-FERMI code. We let $\mu_i$ be shorthand for the set of chemical potentials $\left\{\mu_i^\text{X}\right\}$ if the context specifies a given defect X$^\text{q}$. The quantities $n_d$ and $n_a$ are the concentrations of donor and acceptor charge states, respectively.} 
\label{fig:flowchart2}
\end{figure}

As alluded to above, in order to determine the concentrations of each charge state of the defects under investigation at a given temperature and for a given set of formation energies, we used the SC-FERMI code~\cite{BUCKERIDGE2019329}. The SC-FERMI code reads in as input the density of electronic states $\rho(E)$ (defined as the number of states per energy interval with the energy scale set to zero at the valence band maximum), the band gap $E_g$, the temperature $T$, the lattice vectors $\mathbf{a}_j$ of the unit cell for which $\rho(E)$ is being calculated, and the number of electrons $N_{electron}$ within that unit cell. Additionally, each defect in the system must be provided with a name X and the number of sites $N_{\text{X}}$ on which it can form in the unit cell, which we set equal to 8. Finally, each charge state of a given defect must be provided with the degeneracy $g_{\text{X}^{\text{q}}}$ of the given defect charge state arising from the symmetry of the defect charge state, the formation energy at zero Fermi level $\Delta H_f({\rm X^{\rm q}}, \, \{\mu_i^\text{X}\}, \, 0)$ of the charge state, and an integer q representing the charge state of the defect. The outputs of the SC-FERMI code include $n_0$, $p_0$, $n_{\text{X}^\text{q}}$ for each charge state q, the total concentrations of each defect $n_{\text{X}}$, and the self-consistent Fermi level $E_\text{F}$. The output of the code also includes a charge-transition level diagram, consisting of the lowest formation energy of each defect among the allowed charge states as a function of the energy range within which $\rho(E)$ is defined. In this work, leveraging the SC-FERMI code we use the procedure shown in Fig. \ref{fig:flowchart2}. Due to prohibitive computational cost, in the following analysis the formation energies, unit cell parameters, $\rho(E)$, and $E_g$ are assumed not to change with $T$. 

\subsection{Results and Discussion \label{sec:methodology}}
The calculated band gap of diamond is $E^{\text{DFT}}_{g} = 5.4$~eV in good agreement with other theoretical results~\cite{Deak,Szasz} as well as with experiment~\cite{madelung1991semiconductors}. The relaxed lattice constant was $a^\text{DFT} = 3.54$~\AA~also in good agreement with prior results~\cite{Deak}. We investigated the split-vacancy Sn$V$ defect in diamond in the presence of N$_\text{C}$ in the dilute limit. Thus, we performed two sets of calculations, one for isolated Sn$V$ and the other for isolated N$_\text{C}$. The formation energy we obtain for Sn$V^0$ is $11.45$~eV while for N$^0_{\text{C}}$ we obtain $3.96$~eV. The significantly higher value for the formation energy of neutral Sn$V$ reflects the difficulty associated with inserting an atom of the size of Sn in the diamond lattice. The complete set of formation energies using adiabatic charge-transition levels from the literature~\cite{Kuate2021theor,Gali3} is provided in Table \ref{tab:form_en} as well as the charged-defect degeneracies $g_{\text{X}^{\text{q}}}$. The minimum formation energies are obtained by determining the formation energies for Sn$V$ and N$_\text{C}$ that would lead to neutral defect concentrations typically observed in experiments~\cite{wahl2020direct,kaiser_nitrogen_1959}. The ranges of chemical potentials corresponding to the ranges of formation energies were $\mu_\text{N}\in \left[-11.39,-7.59\right]$ eV/atom and $\mu_{\text{Sn}} \in \left[-4.60,6.58\right]$ eV/atom. The maximum chemical potentials are used to capture the concentrations at which the defects typically occur in experiments, which may be as a result of ion implantation. Ion implantation can locally create very high concentrations of Sn$V$ centers, but phase separation does not occur since these high concentrations are local in nature. The degeneracy factors $g_{\text{X}^{\text{q}}}$ are chosen due to the fact that prior work has shown that the $+1$ state for N$_\text{C}$ has tetrahedral symmetry~\cite{Kuate2021theor}, while the neutral and $-1$ charge states for N$_\text{C}$ have $C_{3v}$ symmetry~\cite{Kuate2021theor,mainwood_nitrogen_1994,kajihara_nitrogen_1991,jones_acceptor_2009,Briddon1992,Lombardi_2003,Cook1966,Ammerlaan1981electron,Smith1959}. Thus, for the $+1$ charge state of N$_\text{C}$ there is only one configuration since all bonds of N to neighboring carbon atoms are equal in length, while for the neutral and $-1$ charge states four configurations can be obtained given that there are four choices for the nearest-neighbor carbon atom that should be associated with the elongated bond. In a similar vein, the Sn$V$ defects have $D_{3d}$ symmetry~\cite{Gali3} and, given a vacant carbon site, there are four ways to choose the adjacent carbon site to leave vacant, such that the Sn atom will lie between the vacant sites.

\begin{table}[ht!]
\caption{Formation energies in eV for various charge states of the Sn$V$ and N$_{\text{C}}$ defects as well as degeneracies.}
\centering
\vspace{1 mm}
\begin{tabular}{ccccc}
\hline\hline 
Defect Species & Charge State & Minimum Formation Energy & Maximum Formation Energy & $g_{\text{X}^{\text{q}}}$\\
\hline
\multirow{5}{*}{Sn$V$}& $2+$ & $-0.93$  & 10.25 & 4\\ 
\cline{2-5}
& $+$ & $-0.56$ & 10.63 & 4\\ 
\cline{2-5}
& 0  & 0.27 & 11.45 & 4\\ 
\cline{2-5}
& $-$  & 2.60 & 13.78 & 4\\ 
\cline{2-5}
& $2-$ & 5.60  & 16.78 & 4\\ 
\hline
\multirow{3}{*}{N$_{\rm C}$} & $+$ & $-3.38$  & 0.42 & 1\\ 
\cline{2-5}
& $0$  & 0.16 & 3.96 & 4\\ 
\cline{2-5}
& $-$  & 4.86 & 8.66  & 4\\ 
\hline
\end{tabular}
\label{tab:form_en}
\end{table}

\subsubsection{The case for only SnV defects in the diamond host\label{sec:methodology}}

For a chosen value of the chemical potential for Sn $\mu_i^{\text{X}} = \mu_{\text{Sn}}$ and temperature $T$, we have computed a self-consistent or equilibrium Fermi level $E_\text{F}^{eq}$ from which we can obtain self-consistent values for the formation energy $\Delta H_f({\rm X^{\rm q}}, \, \{\mu_i^\text{X}\}, \, E_\text{F}^{eq})$, the defect concentrations $n_{\text{X}^\text{q}}(\left\{\mu_i^\text{X}\right\}, T)$, 
the electron concentration $n_0(\left\{\mu_i^\text{X}\right\}, T)$, and the hole concentration $p_0(\left\{\mu_i^\text{X}\right\}, T)$. We could study these self-consistent quantities as a function of $T$ for a particular value of $\mu_\text{Sn}$, but this unfortunately ignores the fact that $\mu_{\text{Sn}}$ also depends on the pressure $P$ and temperature $T$ of the impurity be it in gaseous, liquid, or solid form, which is a function of the growth conditions of Sn in an actual experiment. Here we introduce a model where we assume for simplicity that $\mu_\text{Sn}$ varies linearly along a given sample preparation pathway for illustrative purposes so that we can model possible growth conditions under which impurities such as Sn are incorporated into the host diamond. Our simple model will give us some insight into potential experimental pathways for optimizing the experimental yield of Sn$V$.

The particular trajectory between points in the pathway does not matter as every point is calculated independently of the other points. Every point is an equilibrium calculation that only depends on the $\mu_\text{Sn}$ and $T$ and not on the points that were calculated before it or the points that will be calculated after it. Therefore, though we call the sequence of points a pathway to give some physical intuition for how one might observe, for example, the sequence of Sn$V^-$ yields that are plotted, it is really a series of independent equilibrium calculations. In particular, if an annealing procedure results in some final state that lies within the phase space that we have explored, as long as that final state is an equilibrium state, our results would apply to that state independent of the path by which the system arrived at that state. Similar considerations hold when N$_\text{C}$ is included.

To realize some variation of the chemical potential $\mu_\text{Sn}$ along a given sample preparation pathway, in the actual sample preparation process one can experimentally increase the fluence of Sn atoms during the growth process, which may be through chemical vapor deposition (CVD), or one may apply some other approach such as pressure modulation. We can visualize our model in more detail by constructing a $\mu_{\text{Sn}}-T$-space diagram as illustrated below in Fig. \ref{fig:parameter_square}. Our investigated pathways for the formation of Sn$V^-$ centers begin at one of the green circles and end at one of the red circles in Fig. \ref{fig:parameter_square}. Though we note that the chemical potential $\mu_\text{Sn}$ depends on more experimental parameters than simply the temperature, in our $\mu_{\text{Sn}}-T$ space we can define our pathway as a line  \begin{align}
    \mu_\text{Sn}(T) &= \left(\mu_{\text{Sn}}^{\text{end}}-\mu_{\text{Sn}}^{\text{start}}\right)\frac{T-T^\text{high}}{T^\text{low}-T^\text{high}}+\mu_\text{Sn}^\text{start}
\end{align}
where for for pathway $p_3$ we set $\mu_\text{Sn}^\text{start}$ = $\mu_\text{Sn}^\text{low}$ and 
$\mu_\text{Sn}^\text{end}$ = $\mu_\text{Sn}^\text{high}$;
for pathway $p_{21}$ we set $\mu_\text{Sn}^\text{start}$ = $\mu_\text{Sn}^\text{low}$ and 
$\mu_\text{Sn}^\text{end}$ = $\mu_\text{Sn}^\text{low}$;
for pathway $p_3'$ we set $\mu_\text{Sn}^\text{start}$ = $\mu_\text{Sn}^\text{high}$ and 
$\mu_\text{Sn}^\text{end}$ = $\mu_\text{Sn}^\text{high}$; and for pathway $p_{22}$ we set $\mu_\text{Sn}^\text{start}$ = $\mu_\text{Sn}^\text{high}$ and 
$\mu_\text{Sn}^\text{end}$ = $\mu_\text{Sn}^\text{low}$. The labeling of these pathways is for consistency with the case that includes the presence of N$_\text{C}$.

\begin{figure}[ht!] 
\centering
\includegraphics[width=0.9\textwidth]{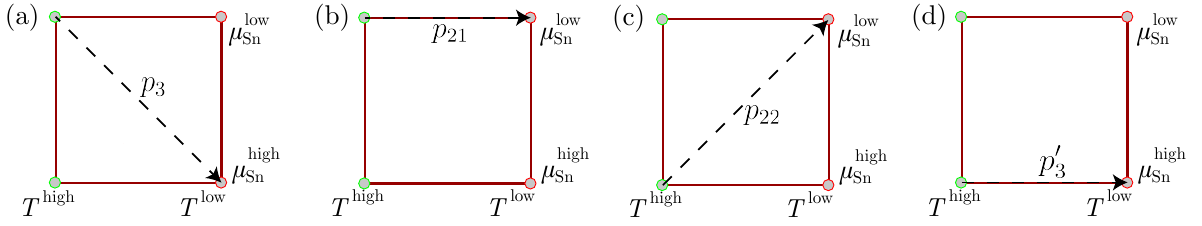}
\caption{Depiction of the parameter space of $T$ and $\mu_\text{Sn}$. The investigated pathways $p_3$, $p_{21}$, $p_{22}$, and $p_3'$, described in the text, are shown in (a), (b), (c) and (d), respectively. The pathways begin at one of the green circles and end at one of the red circles.} 
\label{fig:parameter_square}
\end{figure}

These pathways in $\mu_{\text{Sn}}-T$ space have an important physical interpretation. Equation  (\ref{eq:form_eq}) tells us that for a fixed equilibrium Fermi level, as the chemical potential of Sn increases (decreases) the formation energy of Sn$V$ decreases (increases). In particular, when the chemical potential of Sn is high, $\mu_\text{Sn} = \mu_\text{Sn}^\text{high}$, the formation energy of Sn$V^-$ will be small thus leading to high concentrations for Sn$V^-$. The pathways are labeled according to whether they lead to high or low yields for Sn$V^-$ at low temperature and whether they lead to appreciable or low Sn$V$ concentrations at low temperature. Pathway $p_3$ will lead to high yields of Sn$V^-$, but to a modest Sn$V$ concentration at low temperature. Though they exhibit qualitatively different behavior, pathways $p_{21}$ and $p_{22}$ will both lead to a high yield for Sn$V^-$ at low temperature, but not to an appreciable Sn$V$ concentration at low temperature. Pathway $p_3'$ will lead to an appreciable Sn$V$ concentration at low temperature, but to a vanishingly small yield for Sn$V^-$ at low temperature. 

In order to first consider the case where only Sn$V$ centers are present in diamond, in lowering the temperature we model the ability to increase or decrease the fluence of Sn or the pressure on the sample in an experiment by varying $\mu_\text{Sn}$ so that the formation energies for the charge states of the Sn$V$ defect vary linearly from the minimum values in Table \ref{tab:form_en} to the maximum values in Table \ref{tab:form_en} in 115 steps and the temperature varies from 250~K to 5000~K in 115 steps with each temperature being greater than the previous one by a constant factor. As described above, the lower bounds on the Sn$V$ formation energies are chosen so that the neutral charge state approaches an upper bound on the Sn$V$ concentration typically found in experiment~\cite{wahl2020direct}. Therefore, $n_0(\mu_{\text{Sn}}, T)$, $p_0(\mu_{\text{Sn}}, T)$,  $n_{\text{Sn}V^\text{q}}(\mu_{\text{Sn}}, T)$ for all charge states q, the total concentration of Sn$V$ ($n_{\text{Sn}V}(\mu_{\text{Sn}}, T)$), and $E_\text{F}(\mu_{\text{Sn}}, T) = E_\text{F}^{eq}(\mu_{\text{Sn}}, T)$ were all evaluated at each point in a grid of $116\times116 = 13456$~grid points. Increasing temperature by a constant factor is done to concentrate the sampling at low temperature where we wish to determine the yield. The number of steps for the formation energies is chosen to account for the uncertainty associated with the charge-transition levels, which is about 0.1~eV~\cite{Gali3}. The choice of 5000~K as the upper temperature limit is motivated by the potential for the application of our results in high-pressure high-temperature studies. 

\begin{figure}[ht!] 
\centering
\includegraphics[width=0.99\textwidth]{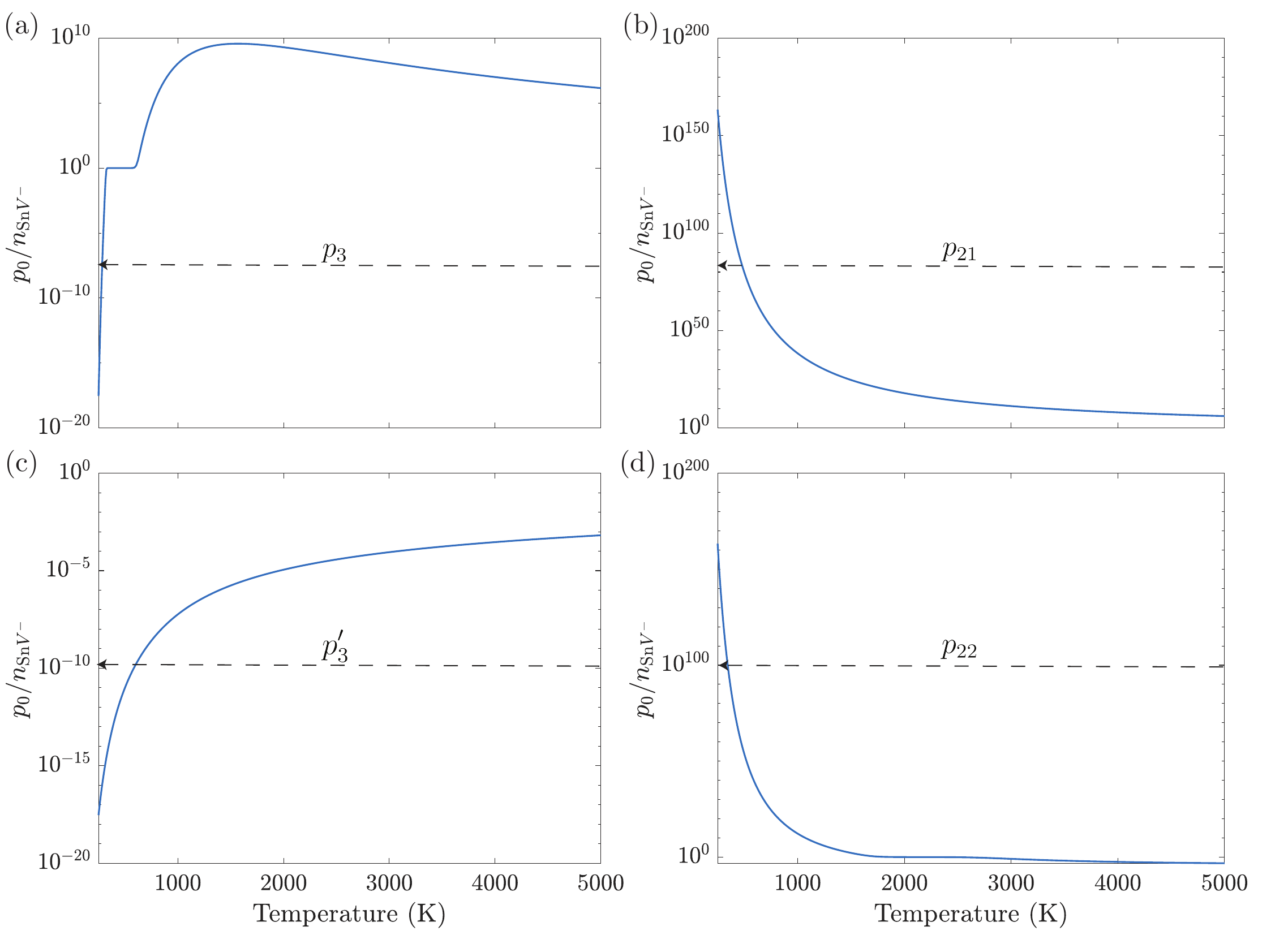}
\caption{The ratio $p_0/n_{\text{Sn}V^-}$ is shown (a) for pathway $p_3$, (b) for pathway $p_{21}$, (c) for pathway $p_3'$, and (d) for pathway $p_{22}$. See Fig. \ref{fig:parameter_square} for the pathways through the parameter space corresponding to $p_3$, $p_{21}$, $p_3'$, and $p_{22}$. All subfigures consider a diamond sample in which only the Sn$V$ is present.} 
\label{fig:p0overSnVm1}
\end{figure}

\begin{figure}[ht!] 
\centering
\includegraphics[width=0.99\textwidth]{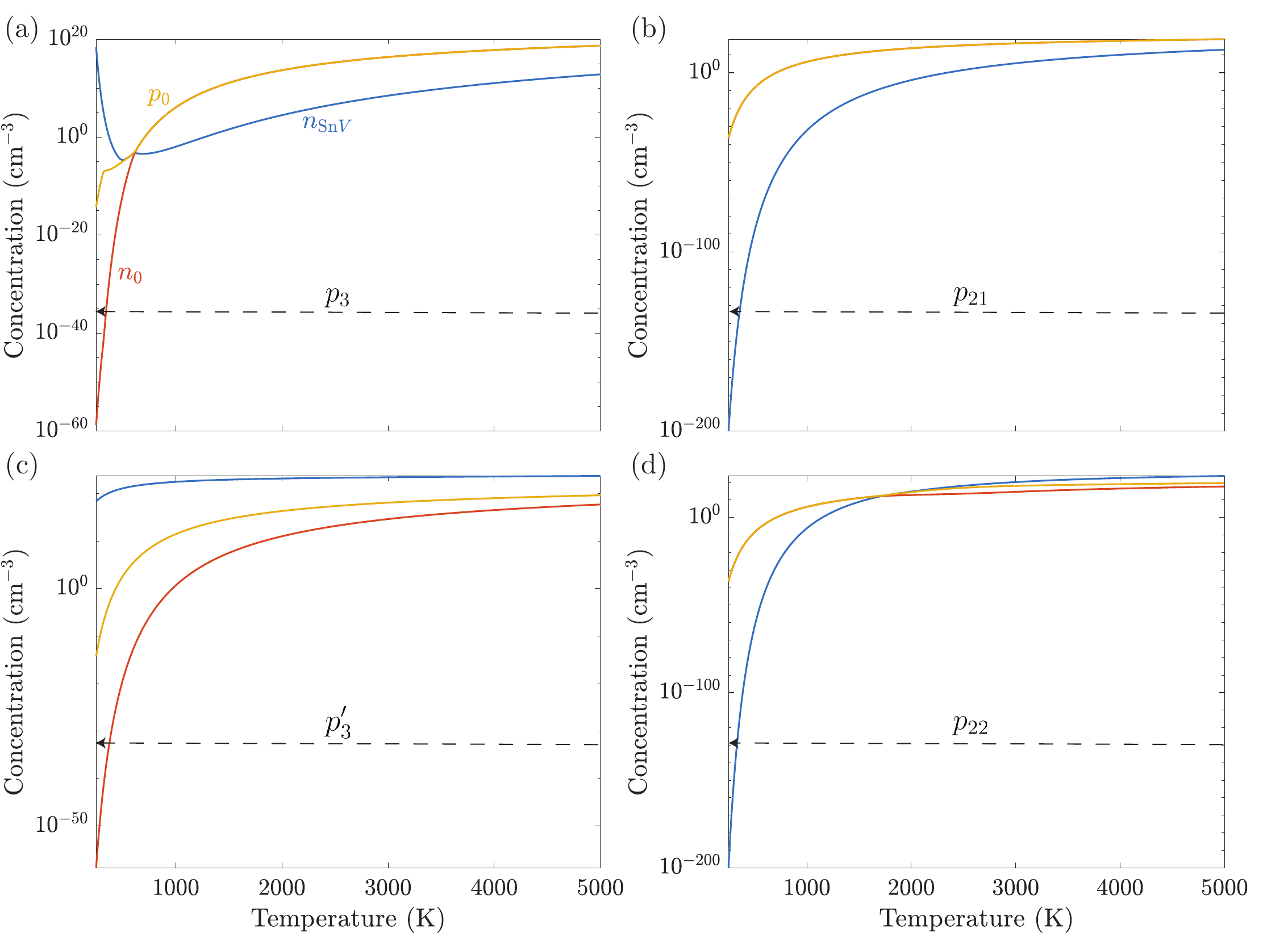}
\caption{The calculated values for $n_0$, $p_0$, and $n_{\text{Sn}V}$ are shown above. The subfigures correspond to results from (a) pathway $p_3$, (b) pathway $p_{21}$, (c) pathway $p_3'$, and (d) pathway $p_{22}$. The pathways $p_3$, $p_{21}$, $p_3'$, and $p_{22}$ are illustrated in Fig. \ref{fig:parameter_square} and we consider a diamond sample in which only the Sn$V$ is present. The curves for $n_0$, $p_0$, and $n_{\text{Sn}V}$ are shown in orange, yellow, and blue, respectively.} 
\label{fig:pathwaysSnVonly_concentrations}
\end{figure}

We first investigate the behavior of the ratio $p_0/n_{\text{Sn}V^-}$, since in the absence of donors the number of holes will limit the yield of Sn$V^-$. If $\mu_\text{Sn}$ is high at low temperature and low at high temperature, then the ratio $p_0/n_{\text{Sn}V^-}$ will pass through the value 1 since at the high temperatures there will be insufficient Sn$V^-$ centers to compensate the holes, which will instead be compensated by electrons, and at low temperature there will be insufficient holes to compensate Sn$V^-$ centers, which will instead be compensated by the positively charged states of the Sn$V$. These results are shown in Fig. \ref{fig:p0overSnVm1} (a) and Fig. \ref{fig:pathwaysSnVonly_concentrations} (a). In the case where $\mu_\text{Sn}$ is always low, there will be insufficient Sn$V^-$ centers to compensate the holes, which will again be compensated by electrons, and the ratio $p_0/n_{\text{Sn}V^-}$ will diverge at low temperature as shown in Fig.  \ref{fig:p0overSnVm1} (b) and Fig. \ref{fig:pathwaysSnVonly_concentrations} (b). As depicted in Fig.  \ref{fig:p0overSnVm1} (c) and Fig. \ref{fig:pathwaysSnVonly_concentrations} (c), in the case where $\mu_\text{Sn}$ is always high, there will be insufficient holes to compensate the Sn$V^-$ centers, which will be compensated by the positively charged states as above, and the ratio $p_0/n_{\text{Sn}V^-}$ will be suppressed at low temperature. Similar to the case where $\mu_\text{Sn}$ is always low, if $\mu_\text{Sn}$ is decreased as the temperature was lowered, then the ratio $p_0/n_{\text{Sn}V^-}$ will diverge as the temperature was lowered because there would be too few Sn$V^-$ centers to compensate every hole (see Fig. \ref{fig:p0overSnVm1} (d) and Fig. \ref{fig:pathwaysSnVonly_concentrations} (d)). 

\begin{figure}[ht!] 
\centering
\includegraphics[width=0.99\textwidth]{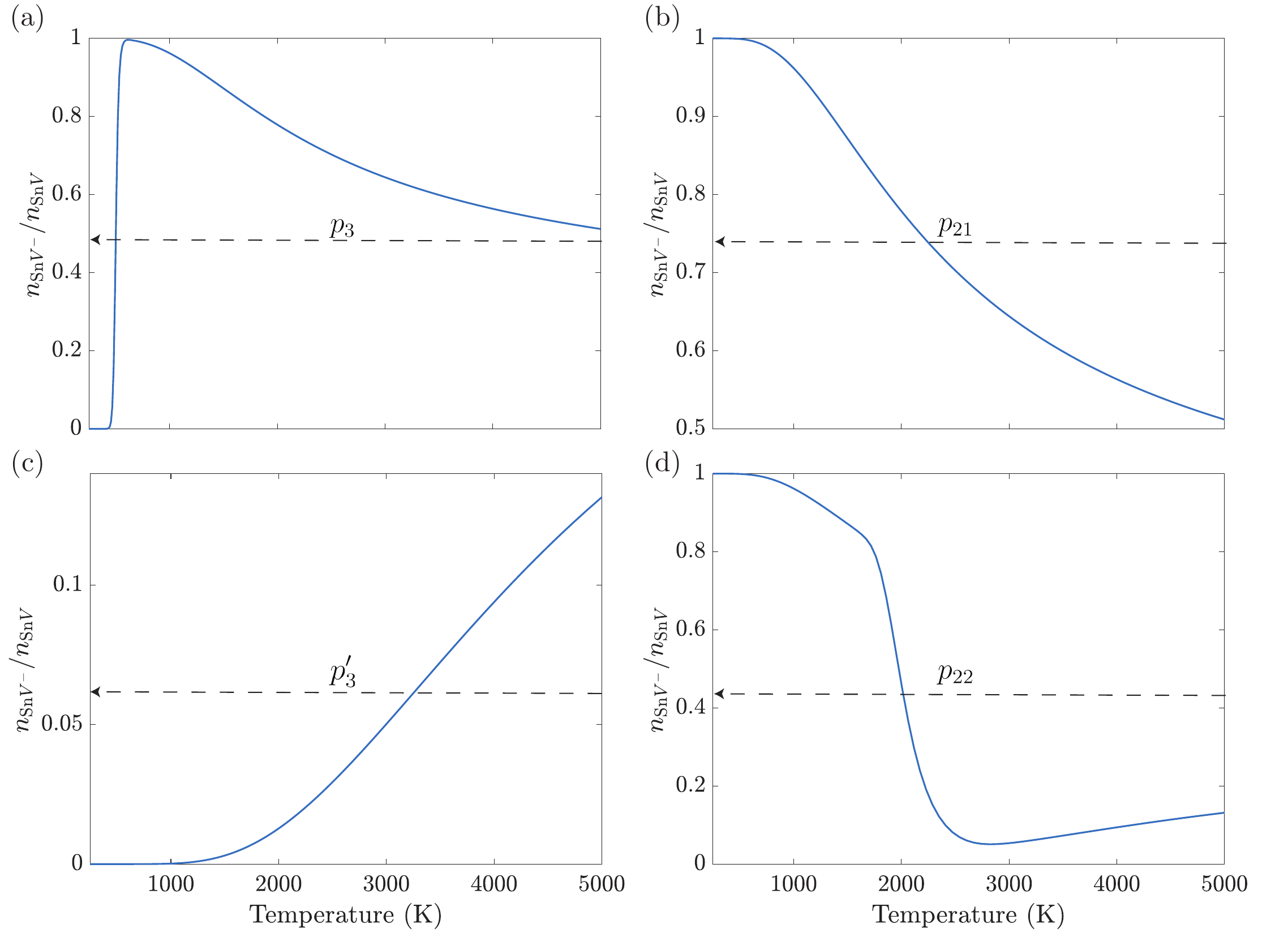}
\caption{The Sn$V^-$ yield is shown in (a) for pathway $p_3$, (b) for pathway $p_{21}$, (c) for pathway $p_3'$, and (d) for pathway $p_{22}$. As above, the pathways through the parameter space corresponding to $p_3$, $p_{21}$, $p_3$, and $p_{22}$ are shown in Fig. \ref{fig:parameter_square} and all subfigures consider a diamond sample in which only the Sn$V$ is present.} 
\label{fig:pathwaysSnVonly_yield}
\end{figure}

We next investigate the concentrations of the species $p_0$, $n_0$, and $n_{\text{Sn}V}$ for the various pathways. Regarding the concentrations of the species $p_0$, $n_0$, and $n_{\text{Sn}V}$ for the pathway $p_3$, the lowering of the temperature suppresses $n_0$, $p_0$, and $n_{\text{Sn}V}$. The concentration $n_{\text{Sn}V}$, however, ultimately increases due to the raising of $\mu_\text{Sn}$ along the pathway (shown in Fig. \ref{fig:pathwaysSnVonly_concentrations} (a)). Given that $p_0$ starts above $n_{\text{Sn}V}$ in the pathway $p_3$, their values must cross at some point along the pathway. It is when they cross that precisely every Sn$V$ is provided with an additional electron by a carrier from a band edge leading to the maximum yield shown in Fig. \ref{fig:pathwaysSnVonly_yield} (a). Beyond that point in the pathway, there are too many Sn$V$ centers to maximize the yield. Before that point in the pathway, it is energetically more favorable for a carrier to travel directly from the valence band to the conduction band so that the yield is again not maximized. The concentrations supporting these conclusions are shown in Fig. \ref{fig:pathwaysSnVonly_concentrations} (a). For the pathway $p_{21}$, $n_{\text{Sn}V}$ is suppressed faster than $n_0$ and $p_0$ as the temperature is decreased. At some point, as shown in Fig. \ref{fig:pathwaysSnVonly_concentrations} (b), $n_{\text{Sn}V}$ becomes so small that the number of carriers required to satisfy every Sn$V$ with an additional electron bears negligible cost leading to the high yield at low temperature that is shown in Fig. \ref{fig:pathwaysSnVonly_yield} (b). For the pathway $p_3'$, Fig. \ref{fig:pathwaysSnVonly_concentrations} (c) illustrates that $n_{\text{Sn}V}$ is suppressed more slowly than $p_0$ implying that there will be insufficient carriers to satisfy every Sn$V$ with an additional electron at sufficiently low temperatures, resulting in the low yields at low temperature shown in Fig. \ref{fig:pathwaysSnVonly_yield} (c). For the pathway $p_{22}$, $n_{\text{Sn}V}$ is again suppressed faster than $n_0$ and $p_0$ as the temperature is decreased. Because $n_{\text{Sn}V}$ is initially at a larger value than $p_0$, at some point the curves corresponding to those concentrations intersect (see Fig. \ref{fig:pathwaysSnVonly_concentrations} (d)) leading to the shoulder in the yield shown in Fig. \ref{fig:pathwaysSnVonly_yield} (d). Beyond the shoulder, as shown in Fig. \ref{fig:pathwaysSnVonly_concentrations} (d), $n_{\text{Sn}V}$ becomes so small that the number of carriers required to satisfy every Sn$V$ with an additional electron bears negligible cost leading to the high yield that is shown in Fig. \ref{fig:pathwaysSnVonly_yield} (d) at low temperature.

\begin{figure}[ht!] 
\centering
\includegraphics[width=0.99\textwidth]{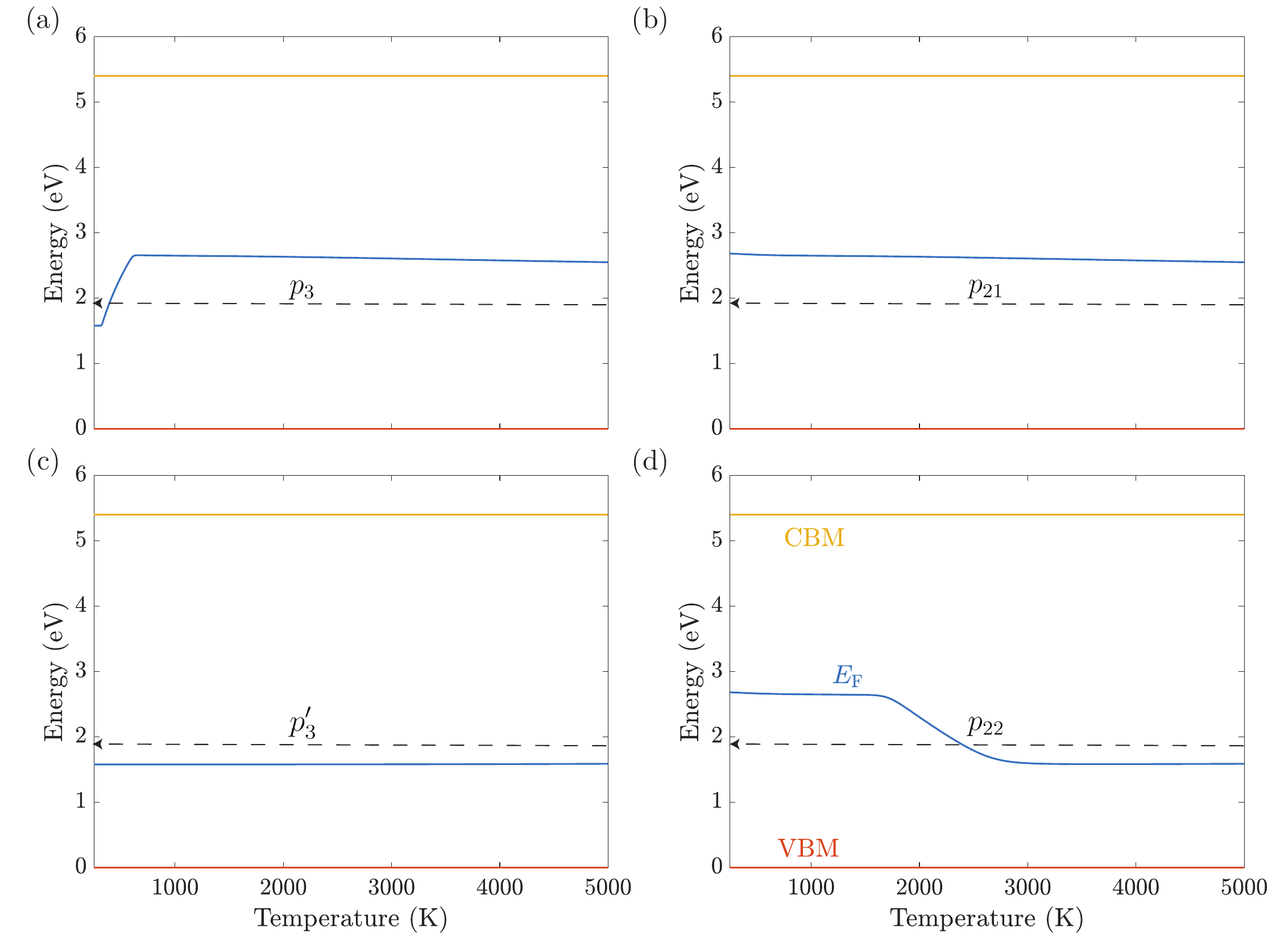}
\caption{The Fermi level, $E_\text{F}$, as a function of temperature is shown in (a) for pathway $p_3$, (b) for pathway $p_{21}$, (c) for pathway $p_3'$, and (d) for pathway $p_{22}$. In Fig. \ref{fig:parameter_square}, the pathways $p_3$, $p_{21}$, $p_3'$, and $p_{22}$ are defined. As in prior figures, only Sn$V$ in the diamond sample is considered. The curves for CBM, $E_\text{F}$, and VBM are shown in yellow, blue, and orange, respectively.} 
\label{fig:pathwaysSnVonly_Fermi}
\end{figure}

We conclude our overview of the various pathways by discussing the behavior of $E_\text{F} = E_\text{F}^{eq}$ for each of the investigated pathways, as shown in Fig. \ref{fig:pathwaysSnVonly_Fermi}. For the pathway $p_3$, at high temperature $n_{\text{Sn}V}$ is small compared to its value at low temperature due to the low chemical potential. Therefore, at high temperature $E_\text{F}$ is approximately equal to the pristine-crystal value of $E_g/2$. As the temperature is lowered, $E_\text{F}$ approaches closer to $E_g/2$ to reflect the suppression of $n_{\text{Sn}V}$ as a result of the lower temperature. When $\mu_\text{Sn}$ and $E_\text{F}$ are such that $n_{\text{Sn}V}$ begins to rise precipitously, $E_\text{F}$ approaches the average of the $(0/+)$ and $(0/-)$ adiabatic charge-transition levels for the Sn$V$, suggesting that the Sn$V$ begins to act as an amphoteric dopant.
These results are depicted in Fig. \ref{fig:pathwaysSnVonly_Fermi} (a). For the pathway $p_{21}$, given the small value of $n_{\text{Sn}V}$ throughout the pathway due to the small value of $\mu_\text{Sn}$, $E_\text{F}$ is pinned near the pristine-crystal value of $E_g/2$ (see Fig. \ref{fig:pathwaysSnVonly_Fermi} (b)). For the pathway $p_3'$, $n_{\text{Sn}V}$ is always high due to a high $\mu_\text{Sn}$. Therefore, $E_\text{F}$ reflects the abundance of Sn$V$ through the fact that $E_\text{F}$ is pinned near the average of the $(0/+)$ and $(0/-)$ adiabatic charge-transition levels for the Sn$V$, again suggesting that the Sn$V$ is acting as an amphoteric dopant (see Fig. \ref{fig:pathwaysSnVonly_Fermi} (c)). For the case of the pathway $p_{22}$, $n_{\text{Sn}V}$ is initially large due to the large value of $\mu_\text{Sn}$. As a result, $E_\text{F}$ approaches the average of the $(0/+)$ and $(0/-)$ adiabatic charge-transition levels for the Sn$V$, suggesting that the Sn$V$ initially acts as an amphoteric dopant. For the remainder of the pathway the suppression of $n_{\text{Sn}V}$ due both to temperature and to a low $\mu_\text{Sn}$ leads to $E_\text{F}$ approaching the pristine-crystal value of $E_g/2$ (see Fig. \ref{fig:pathwaysSnVonly_Fermi} (d)). 

\subsubsection{The case for both SnV and N defects in the diamond host \label{sec:methodology}}

\begin{figure}[ht!] 
\centering
\includegraphics[width=0.4\textwidth]{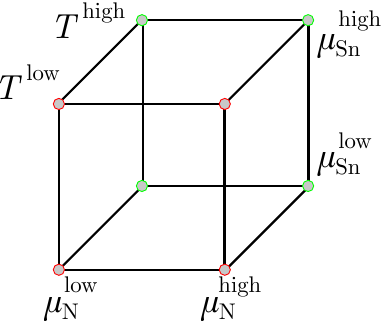}
\caption{Depiction of the parameter space of $T$, $\mu_\text{Sn}$, and $\mu_\text{N}$. Investigated pathways begin at one of the green circles and end at one of the red circles.} 
\label{fig:parameter_cube}
\end{figure}

We next introduce N$_\text{C}$ into the system and model the ability to increase or decrease the fluence of Sn or N in an experiment by varying the chemical potentials of N and Sn, $\mu_\text{N}$ and $\mu_\text{Sn}$. Again, the formation energies for the Sn$V$ defects vary linearly between the minimum and maximum values in Table \ref{tab:form_en} in 115 steps and the temperature varies from 250~K to 5000~K in 115 steps as well, with each subsequent temperature being greater than the previous temperature by a constant factor as above. For the various charge states of the N$_{\text{C}}$ defects, 40 steps separate their minimum formation energies from their maximum formation energies in Table \ref{tab:form_en}. Similarly to the case of only the Sn$V$, the minimum formation energy was chosen so that the concentration of neutral N$_\text{C}$ would replicate high concentrations typically found in experiments~\cite{kaiser_nitrogen_1959}. Therefore, $n_0(\mu_{\text{Sn}},\mu_{\text{N}}, T)$, $p_0(\mu_{\text{Sn}}, \mu_{\text{N}},T)$,  $n_{\text{Sn}V^\text{q}}(\mu_{\text{Sn}},\mu_{\text{N}}, T)$ for all charge states q, $n_{\text{N}_\text{C}^\text{q}}(\mu_{\text{Sn}},\mu_{\text{N}}, T)$ for all charge states q, the total concentration of Sn$V$ ($n_{\text{Sn}V}(\mu_{\text{Sn}},\mu_{\text{N}}, T)$), the total concentration of $\text{N}_\text{C}$ ($n_{\text{N}_\text{C}}(\mu_{\text{Sn}},\mu_{\text{N}}, T)$), and $E_\text{F}(\mu_{\text{Sn}},\mu_{\text{N}}, T)$  were computed at each point in a grid of $ 116\times116\times41 = 551696$~grid points. Similar to the case of the Sn$V$, the number of steps for the formation energies of the N$_\text{C}$ species was chosen to account for the uncertainty associated with the charge-transition levels, which is about 0.1~eV~\cite{Kuate2021theor}. The parameter space is therefore three-dimensional parameter space, with $\mu_\text{Sn}$, $T$, and $\mu_\text{N}$ labeling the dimensions. 

\begin{figure}[ht!] 
\centering
\includegraphics[width=0.5\textwidth]{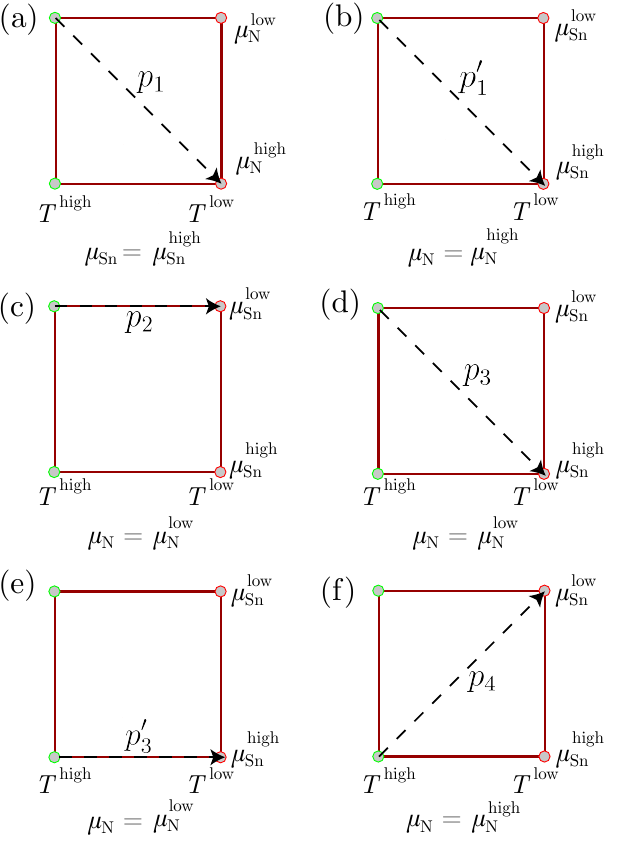}
\caption{Depiction of slices of the parameter space of $T$, $\mu_\text{Sn}$, and $\mu_\text{N}$. The investigated pathways $p_1$, $p_{1}'$, $p_{2}$, $p_3$, $p_3'$, and $p_4$, described in the text, are shown in (a), (b), (c), (d), (e), and (f), respectively. The pathways begin at one of the green circles and end at one of the red circles.}
\label{fig:parameter_cube_pathways}
\end{figure}

We find six classes of pathways with prototypical pathways labeled by $p_1$, $p_1'$, $p_2$, $p_3$, $p_3'$, and $p_4$.  As above, the pathways are labeled according to whether they lead to high or low yields for Sn$V^-$ at low temperature and whether they lead to appreciable or low Sn$V$ concentrations at low temperature. The class of pathways labeled by $p_1$ will lead both to high yields of Sn$V^-$ and to an appreciable Sn$V$ concentration at low temperature. The class of pathways labeled by $p_1'$ will lead to modest yields except at the very lowest temperature where high yields can be obtained, but the concentration of Sn$V$ will be appreciable at low temperature. The class of pathways labeled by $p_2$ will lead to a high yield for Sn$V^-$ at low temperature, but not to an appreciable Sn$V$ concentration at low temperature, as above. The class of pathways labeled by $p_3$ will lead to a modest Sn$V$ concentration at low temperature, but to a high yield for Sn$V^-$ at low temperature, similar again to the results above. Also as discussed earlier, the class of pathways labeled by $p_3'$ will lead to an appreciable Sn$V$ concentration, but to a vanishingly small yield of Sn$V^-$ at low temperature. The class of pathways labeled by $p_4$ will lead to a low yield for Sn$V^-$ at low temperature and to no appreciable Sn$V$ concentration at low temperature. The prototypical pathways for the six classes of pathways are in Fig. \ref{fig:parameter_cube_pathways}.

\begin{figure}[ht!] 
\centering
\includegraphics[width=0.99\textwidth]{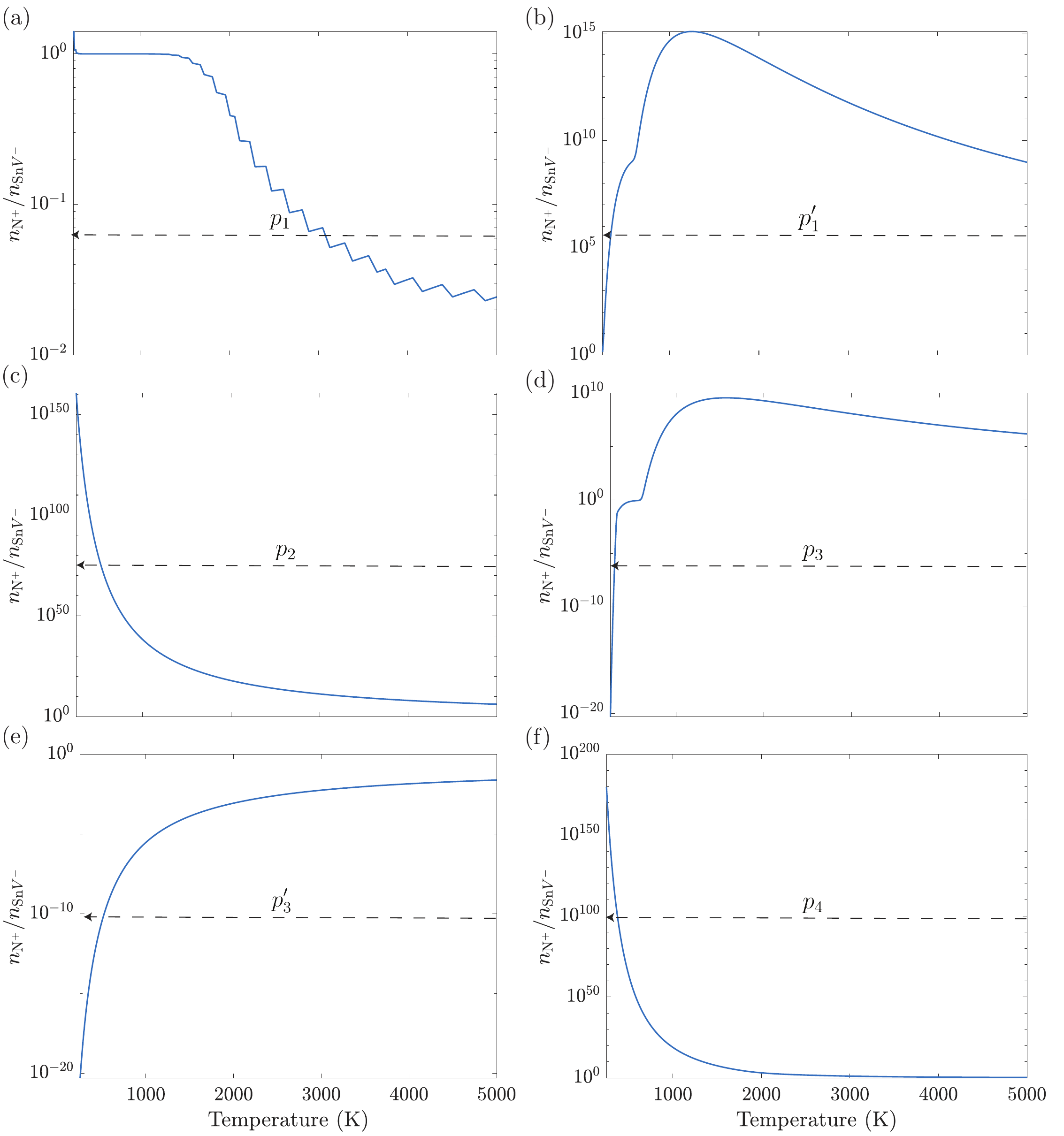}
\caption{The ratio $n_{\text{N}^+}/n_{\text{Sn}V^-} = n_{\text{N}_\text{C}^+}/n_{\text{Sn}V^-}$ is shown (a) for pathway $p_1$, (b) for pathway $p_1'$, (c) for pathway $p_2$, (d) for pathway $p_3$, (e) for pathway $p_3'$, and (f) for pathway $p_4$. See Fig. \ref{fig:parameter_cube_pathways} for the pathways through the parameter space corresponding to $p_1$, $p_1'$, $p_2$, $p_3$, $p_3'$, and $p_4$. All subfigures consider a diamond sample in which Sn$V$ and ${\text{N}_\text{C}}$ are present.} 
\label{fig:Np1overSnVm1}
\end{figure}

\begin{figure}[ht!] 
\centering
\includegraphics[width=0.99\textwidth]{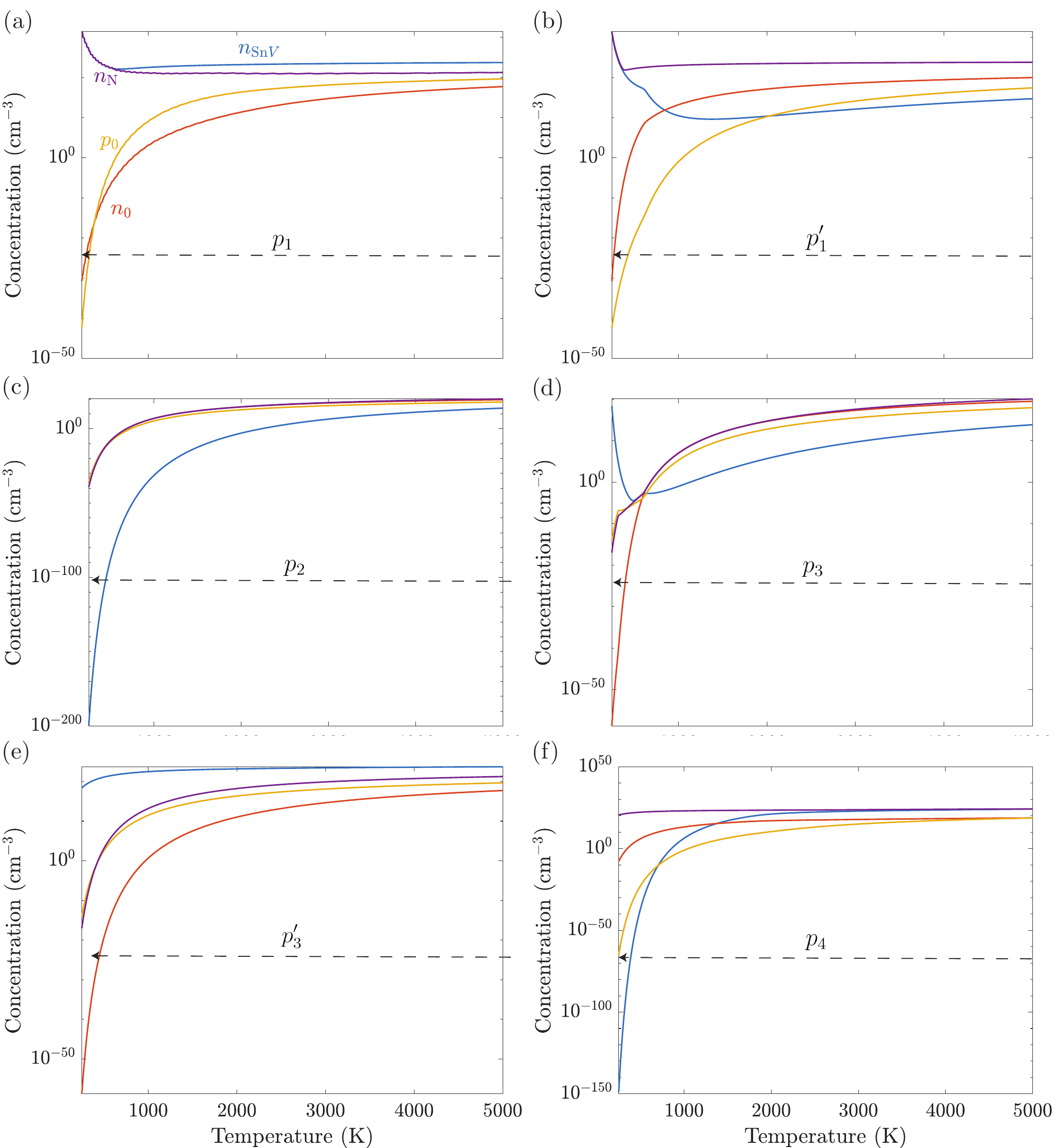}
\caption{The calculated values for $n_0$, $p_0$, $n_{\text{Sn}V}$, and $n_{\text{N}}$ are shown above. The subfigures correspond to results from (a) pathway $p_1$, (b) pathway $p_1'$, (c) pathway $p_2$, (d) pathway $p_3$, (e) pathway $p_3'$, and (f) pathway $p_4$. The precise pathways through the parameter space corresponding to $p_1$, $p_1'$, $p_2$, $p_3$, $p_3'$, and $p_4$ are depicted in Fig. \ref{fig:parameter_cube_pathways}. Again, all subfigures consider a diamond sample in which both Sn$V$ and N$_\text{C}$ are present. The curves for $n_0$, $p_0$, $n_{\text{Sn}V}$, and $n_{\text{N}}$ are shown in orange, yellow, and blue, and purple, respectively.} 
\label{fig:pathwaysSnVandN_concentrations}
\end{figure}

For Sn$V$ centers in the presence of N$_\text{C}$ in diamond, we first investigate the behavior of the ratio $n_{\text{N}^+}/n_{\text{Sn}V^-} = n_{\text{N}^+_\text{C}}/n_{\text{Sn}V^-} $ since the N$_\text{C}$ donors will limit the yield of Sn$V^-$. If $\mu_\text{Sn}$ is high at low temperature and $\mu_\text{N}$ increases throughout the pathway as shown in Figs. \ref{fig:Np1overSnVm1} (a), then the ratio $n_{\text{N}^+}/n_{\text{Sn}V^-}$ approaches 1 as the temperature is lowered since the number of N$_\text{C}^+$ approaches the quantity needed to compensate Sn$V^-$ as the pathway progresses. The total defect concentrations supporting this argument are shown in Fig. \ref{fig:pathwaysSnVandN_concentrations} (a). If $\mu_\text{Sn}$ is increased throughout the pathway and $\mu_\text{N}$ is high throughout the pathway, then the concentration of N$_\text{C}$ will be high throughout the pathway while the concentration of Sn$V$ will eventually rise to approach the concentration of N$_\text{C}$ as the temperature is lowered (as shown in Fig. \ref{fig:pathwaysSnVandN_concentrations} (b)). Thus, at low temperature the ratio $n_{\text{N}^+}/n_{\text{Sn}V^-}$ will approach 1, as shown in Fig. \ref{fig:Np1overSnVm1} (b). If $\mu_\text{Sn}$ and $\mu_\text{N}$ are both low throughout the pathway, then, since the concentration of N$_\text{C}$ is suppressed more slowly than the concentration of Sn$V$ (as shown in Fig. \ref{fig:pathwaysSnVandN_concentrations} (c)), the ratio $n_{\text{N}^+}/n_{\text{Sn}V^-}$ will diverge, as shown in Fig. \ref{fig:Np1overSnVm1} (c). Similar to the arguments above, if $\mu_\text{Sn}$ increases throughout the pathway and $\mu_\text{N}$ is low throughout the pathway then the ratio $n_{\text{N}^+}/n_{\text{Sn}V^-}$ will pass through the value 1 since at the high temperatures there will be insufficient Sn$V^-$ centers to compensate the N$_\text{C}^+$, which will instead be compensated by electrons, and at low temperature there will be insufficient N$_\text{C}^+$ to compensate Sn$V^-$ centers, which will instead be compensated by the positively charged states of the Sn$V$. We show these results in Fig. \ref{fig:Np1overSnVm1} (d) and Fig. \ref{fig:pathwaysSnVandN_concentrations} (d). In the case where $\mu_\text{Sn}$ is always high and $\mu_\text{N}$ is low, there will be insufficient N$_\text{C}^+$ to compensate the Sn$V^-$, which will be compensated by the positively charged states of the Sn$V$ so that the ratio $n_{\text{N}^+}/n_{\text{Sn}V^-}$ will be suppressed. See Fig. \ref{fig:Np1overSnVm1} (e) and Fig. \ref{fig:pathwaysSnVandN_concentrations} (e) for these results. Finally, in the case where $\mu_\text{Sn}$ decreases and $\mu_\text{N}$ is always high, there will be insufficient Sn$V^-$ centers to compensate the N$_\text{C}^+$, which will be compensated by the negatively charged state of the N$_\text{C}$ defect, and the ratio $n_{\text{N}^+}/n_{\text{Sn}V^-}$ will diverge at low temperature as shown in Fig.  \ref{fig:Np1overSnVm1} (f) and Fig. \ref{fig:pathwaysSnVandN_concentrations} (f).

\begin{figure}[ht!] 
\centering
\includegraphics[width=0.99\textwidth]{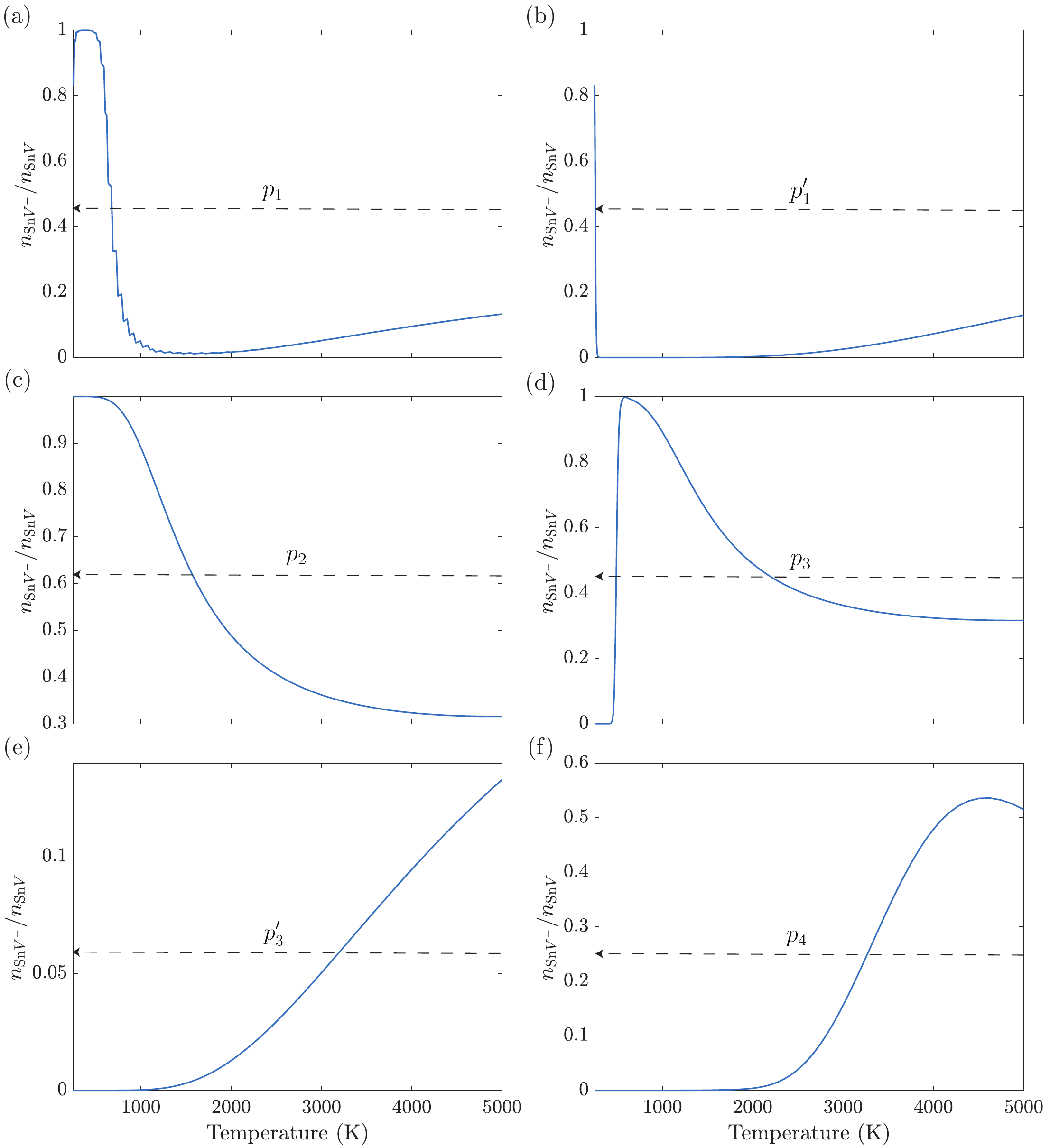}
\caption{The Sn$V^-$ yield is shown in (a) for pathway $p_1$, (b) for pathway $p_1'$, (c) for pathway $p_2$, (d) for pathway $p_3$, (e) for pathway $p_3'$, and (f) for pathway $p_4$. The pathways $p_1$, $p_1'$, $p_2$, $p_3$, $p_3'$, and $p_4$ are illustrated in Fig. \ref{fig:parameter_cube_pathways}. As above, all subfigures consider a diamond sample in which both Sn$V$ and N$_\text{C}$ are present.} 
\label{fig:pathwaysSnVandN_yield}
\end{figure}

We now turn to an investigation of the concentrations of the species $p_0$, $n_0$, $n_{\text{Sn}V}$, and $n_\text{N} = n_{\text{N}_\text{C}}$ for the various pathways. For the prototypical pathway $p_1$ where $\mu_\text{N}$ is raised throughout the pathway and $\mu_\text{Sn}$ is high throughout the pathway, the $n_\text{N}$ increases to approach $n_{\text{Sn}V}$ as shown in Fig. \ref{fig:pathwaysSnVandN_concentrations} (a). When the concentrations are equal, the yield is maximized as shown in Fig. \ref{fig:pathwaysSnVandN_yield} (a). For the prototypical pathway $p_1'$ where $\mu_\text{N}$ is high throughout the pathway and $\mu_\text{Sn}$ is raised throughout the pathway, $n_\text{N}$ is higher than $n_{\text{Sn}V}$ throughout the majority of the pathway (see Fig. \ref{fig:pathwaysSnVandN_concentrations} (b)). The N$_\text{C}^+$ point defects are mostly compensated by negative charge states of the same defect when $n_\text{N}$ is higher than $n_{\text{Sn}V}$ and holes are mostly compensated by electrons leading to the low yields at high temperature shown in Fig. \ref{fig:pathwaysSnVandN_yield} (b). At sufficiently low temperature, $n_{\text{Sn}V}$ approaches the value of $n_\text{N}$ and the yield of Sn$V^-$ approaches unity. For the prototypical pathway $p_2$ where $\mu_{\text{Sn}}$ and $\mu_\text{N}$ are low throughout the pathway, $n_{\text{Sn}V}$ is suppressed faster than $n_0$, $p_0$, and $n_\text{N}$ as the temperature is decreased as shown in Fig. \ref{fig:pathwaysSnVandN_concentrations} (c). In that pathway, there are sufficiently many donors, $\text{N}_\text{C}$, to ionize given the high chemical potential of $\mu_\text{N}$ so that every Sn$V$ can be provided with an electron, which results in the high yield of Sn$V^-$ shown in Fig. \ref{fig:pathwaysSnVandN_yield} (c) at low temperature. In the prototypical pathway $p_3$, similar to the results above, the lowering of the temperature suppresses $n_0$, $p_0$, $n_{\text{Sn}V}$ and $n_\text{N}$. The concentration $n_{\text{Sn}V}$, however, ultimately increases due to the raising of $\mu_\text{Sn}$ along the pathway (shown in Fig. \ref{fig:pathwaysSnVandN_concentrations} (d)) resulting in a region of high yield when the values of $n_{\text{Sn}V}$ and $n_\text{N}$ intersect (see Fig. \ref{fig:pathwaysSnVandN_yield} (d)). In the case of this prototypical pathway, throughout $\mu_\text{N}$ is kept low. Again similar to the results above, when $\mu_{\text{Sn}}$ is high throughout a pathway and $\mu_{\text{N}}$ is low throughout a pathway, as in the prototypical pathway indicated by $p_3'$, the concentration of Sn$V$ centers is always much higher than that of either N$_\text{C}$ or holes (see Fig. \ref{fig:pathwaysSnVandN_concentrations} (e)), so that the yield of Sn$V^-$ is suppressed at low temperature (see Fig. \ref{fig:pathwaysSnVandN_yield} (e)). The prototypical pathway $p_4$, where $\mu_\text{N}$ is kept high throughout the pathway and $\mu_\text{Sn}$ is lowered throughout the pathway leads to a suppressed concentration of Sn$V$ centers at low temperature as shown in Fig. \ref{fig:pathwaysSnVandN_concentrations} (f). Given that $n_\text{N}$ is increasing relative to $n_{\text{Sn}V}$ and $p_0$ is decreasing relative to $n_0$ in that pathway, the maximum yield occurs when the availability of holes and N$_\text{C}$ donors is maximized (see Fig. \ref{fig:pathwaysSnVandN_yield} (f)). Beyond that point, N$_\text{C}^+$ prefer to be compensated by electrons or by negative charge states of the same defect.  

\begin{figure}[ht!] 
\centering
\includegraphics[width=0.99\textwidth]{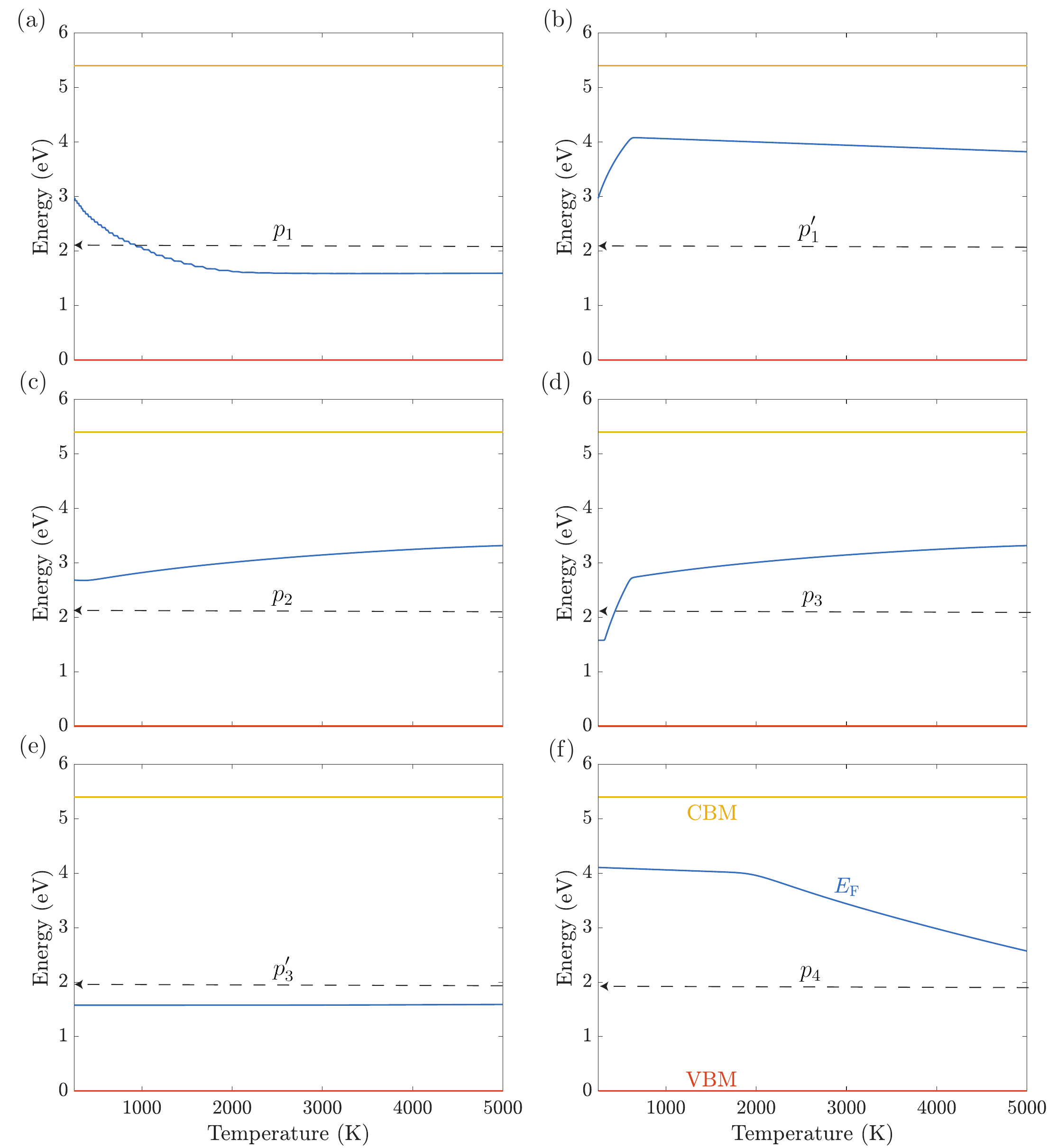}
\caption{The Fermi level, $E_\text{F}$, as a function of temperature is shown in (a) for pathway $p_1$, (b) for pathway $p_1'$, (c) for pathway $p_2$, (d) for pathway $p_3$, (e) for pathway $p_3'$, and (f) for pathway $p_4$. The depiction of the pathways $p_1$, $p_1'$, $p_2$, $p_3$, $p_3'$, and $p_4$ is provided in Fig. \ref{fig:parameter_cube_pathways}. In the subfigures, both Sn$V$ and N$_\text{C}$ are assumed to be present in the diamond sample. The curves for CBM, $E_\text{F}$, and VBM are shown in yellow, blue, and orange, respectively.} 
\label{fig:pathwaysSnVandN_Fermi}
\end{figure}

Finally, as above, we end with a discussion of the behavior of $E_\text{F} = E_\text{F}^{eq}$, as illustrated in Fig. \ref{fig:pathwaysSnVandN_Fermi}. For the prototypical pathway $p_1$, $n_{\text{Sn}V}$ is initially large so that the Fermi level is pinned near the average of the $(0/+)$ and $(0/-)$ adiabatic charge-transition levels for Sn$V$ in diamond, reflecting the fact that Sn$V$ is acting as an amphoteric dopant. As the temperature is lowered, $E_\text{F}$ approaches the average of the $(0/+)$ adiabatic transition level for N$_\text{C}$ and the $(0/-)$ adiabatic transition level for the Sn$V$ to reflect the fact that the dopants are present in significant concentrations and in approximately equal proportion. These results are shown in Fig. \ref{fig:pathwaysSnVandN_Fermi} (a). For the prototypical pathway $p_1'$, $n_{\text{N}}$ is initially much larger than the concentrations of other species so that N$_\text{C}^+$ are compensated by N$_\text{C}^-$ and $E_\text{F}$ is near the $(0/+)$ and $(0/-)$ adiabatic charge-transition levels for the N$_\text{C}$ in diamond. As the temperature is lowered, $n_{\text{Sn}V}$ increases to the value of $n_{\text{N}}$ so that $E_\text{F}$ approaches the value of the average of the $(0/+)$ adiabatic transition level for N$_\text{C}$ and the $(0/-)$ adiabatic transition level for the Sn$V$. For the prototypical pathway $p_2$, the most abundant species are initially N$_\text{C}$ and electrons so that $E_\text{F}$ is initially near the $(0/+)$ adiabatic charge-transition for the N$_\text{C}$. As the temperature is lowered, the concentrations of the various species are suppressed so that $E_\text{F}$ approaches the value $E_g/2$. The depiction of these results is provided in Fig. \ref{fig:pathwaysSnVandN_Fermi} (b). For the prototypical pathway $p_3$, N$_\text{C}$ and electrons are again initially the most abundant species so that $E_\text{F}$ is initially near the $(0/+)$ adiabatic charge-transition for the N$_\text{C}$. At some point in the pathway, $E_\text{F}$ and $\mu_\text{Sn}$ are such that $n_{\text{Sn}V}$, begins to rise precipitously. This rise is accompanied by a drop in $E_\text{F}$ so that it approaches the average of the $(0/+)$ and $(0/-)$ adiabatic charge-transition levels for the Sn$V$, which implies as above that the Sn$V$ begins to act as an amphoteric dopant. The corresponding results can be found in Fig. \ref{fig:pathwaysSnVandN_Fermi} (c). Between the two ends of the pathway, $E_\text{F}$ approaches $E_g/2$ to reflect the suppression of the various species. For the prototypical pathway $p_3'$, as shown in Fig.  \ref{fig:pathwaysSnVandN_Fermi} (d), $n_{\text{Sn}V}$ is appreciable throughout the pathway so that $E_\text{F}$ is always pinned near the average of the $(0/+)$ and $(0/-)$ adiabatic charge-transition levels for the Sn$V$. For the prototypical pathway $p_4$, the value of $E_\text{F}$ is initially near a value that reflects the presence not only of Sn$V^-$ and N$_\text{C}^+$, but also of other charge states for Sn$V$ centers and for N$_\text{C}$ defects. As the temperature is lowered, various other species are suppressed to a greater extent so that the system is left with an abundance of N$_\text{C}$ defects. At low temperature therefore, we observe that $E_\text{F}$ approaches the average of the $(0/+)$ and $(0/-)$ adiabatic charge-transition levels for the N$_\text{C}$ in diamond (see Fig. \ref{fig:pathwaysSnVandN_Fermi} (e)). 

We note that we can cast the behavior of the Fermi level within the framework we had proposed in earlier work~\cite{kuate_investigating_2024}. Essentially, if a single charge state of a defect dominates, the Fermi level will be pinned near the charge transition level for that charge state to reflect exchange of charge between the band edges and the defect in that charge state, with the donor or acceptor level equal to the charge transition level for the charge state. When two charge states dominate, the Fermi level is pinned near the average of the corresponding charge transition levels to reflect exchange of charge between the band edges and the effective defect consisting of the two charged defect states, with an effective donor or acceptor level given by the average of the corresponding charge transition levels. For more than two defects, the effective defect level must be obtained through the general self-consistent approach outlined above and charge is exchanged between the band edges and the effective defect with the corresponding effective defect level.

\begin{figure}[ht!] 
\centering
\includegraphics[width=0.99\textwidth]{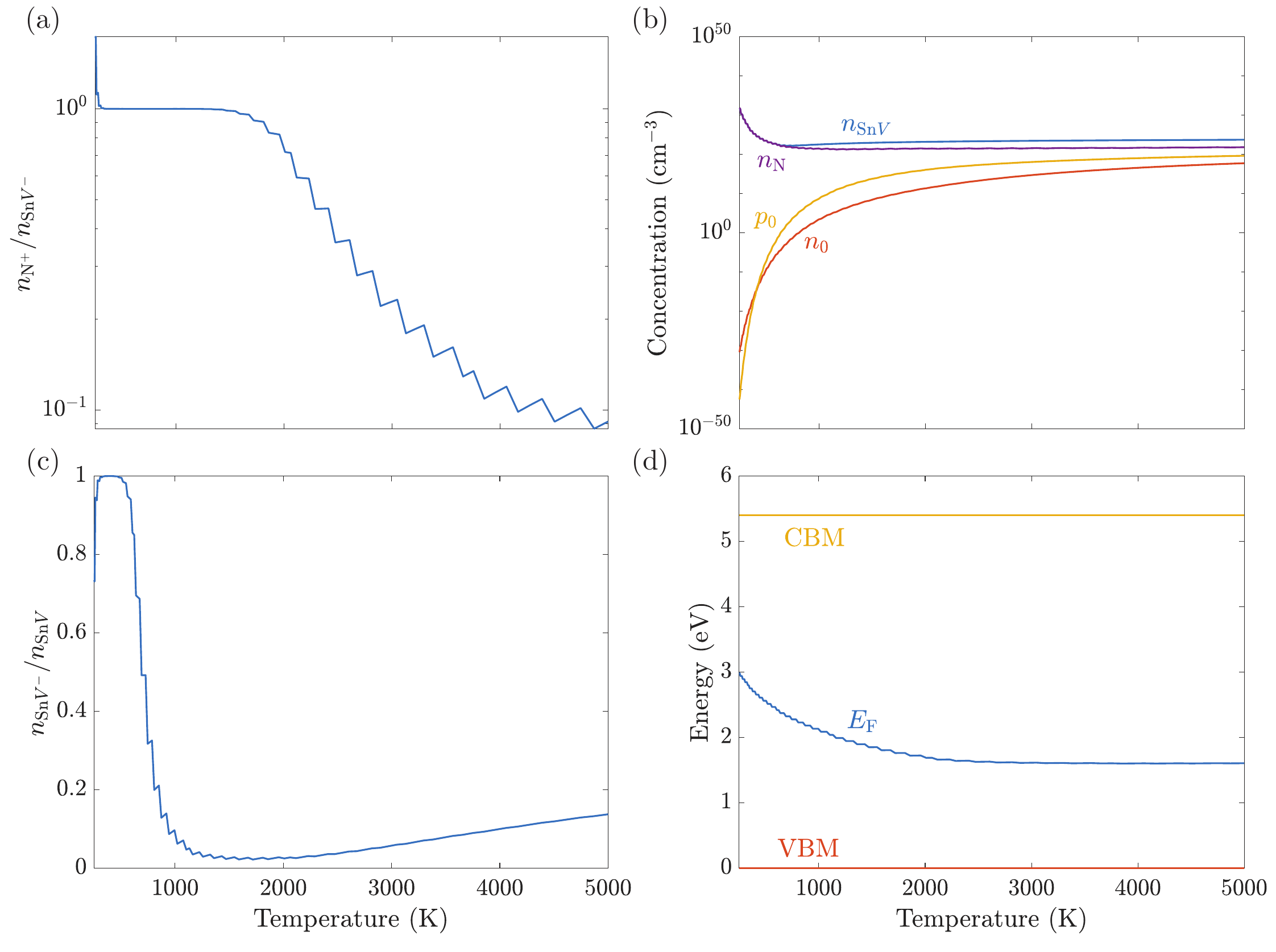}
\caption{The ratio $n_{\text{N}^+}/n_{\text{Sn}V^-} = n_{\text{N}_\text{C}^+}/n_{\text{Sn}V^-}$; calculated values for $n_0$, $p_0$, $n_{\text{Sn}V}$, and $n_{\text{N}}$; Sn$V^-$ yield; and Fermi level, $E_\text{F}$, as a function of temperature are shown in (a), (b), (c), and (d), respectively, for the pathway $p_1$ with $g_{{\text{N}^{\text{+}}_\text{C}}} = 4$. The depiction of the pathway $p_1$ is provided in Fig. \ref{fig:parameter_cube_pathways}. As in Figs. \ref{fig:Np1overSnVm1}-\ref{fig:pathwaysSnVandN_Fermi}, both Sn$V$ and N$_\text{C}$ are assumed to be present in the diamond sample. The curves for $n_0$, $p_0$, $n_{\text{Sn}V}$, and $n_{\text{N}}$ in (b) are shown in orange, yellow, and blue, and purple, respectively. The curves for CBM, $E_\text{F}$, and VBM in (d) are shown in yellow, blue, and orange, respectively.} 
\label{fig:checkNp14}
\end{figure}

We also comment on the degeneracy factor, $g_{\text{X}^{\text{q}}}$, that is used to account for the symmetry of a given charge state q of a defect X. Prior work has shown that equivalent electronic configurations can arise depending on the degeneracy of the band edges, a phenomenon which is dependent on temperature~\cite{ma_carrier_2011,mutter_calculation_2015}. We therefore investigate the effect of changing the degeneracy factor for the $+1$ charge state of N$_\text{C}$ from $g_{{\text{N}^{\text{+}}_\text{C}}} = 1$ to $g_{{\text{N}^{\text{+}}_\text{C}}} = 4$ in the pathway $p_1$. The results are depicted in Fig. \ref{fig:checkNp14}. We find negligible quantitative change in the results and no qualitative change in the results. Given that the concentrations in Eq. (\ref{eq:model-1}) depend linearly on the degeneracy while they depend exponentially on the temperature and the chemical potentials, we indeed do not expect appreciable changes in the results.  

\begin{figure}[ht!] 
\centering
\includegraphics[width=0.9\textwidth]{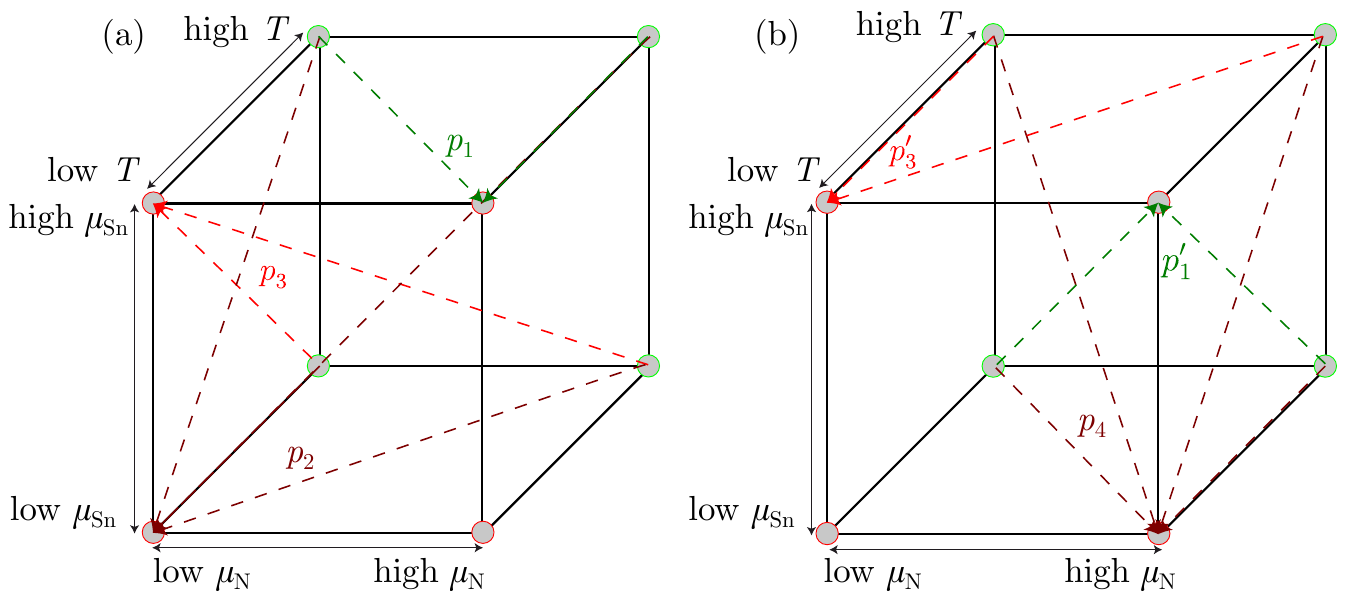}
\caption{The classes of pathways leading to near unity yields for Sn$V^-$ at low temperature are shown in (a) and the classes of pathways that do not lead to near unity yields for Sn$V^-$ at low temperature are shown in (b). The class of pathways labeled by $p_1$ leads both to a high yield and to an appreciable defect concentration for Sn$V^-$, while the classes of pathways labeled by $p_1'$, $p_2$, $p_3$, $p_3'$, and $p_4$ do not.} 
\label{fig:pathwaysall}
\end{figure}

The results indicate that not only is it important to have high propensity for the formation of Sn$V$ centers and N$_\text{C}$ defects to obtain high yields of Sn$V^-$ centers, it is also important to maintain the high propensity for the formation of Sn$V$ centers throughout the preparation pathway. We summarize the classes of pathways in Fig. \ref{fig:pathwaysall}. Though the linear dependence of the chemical potential on temperature is arbitrary, the results nonetheless demonstrate that there is a set of parameter values for which both high yields and appreciable concentrations can be observed for the Sn$V^-$ at low temperature. Since each calculated point is independent of the others, the path by which the appropriate set of parameter values is arrived at is not important. What is important is the determination of some continuous path through the parameter space that brings the diamond sample from the initial conditions of the experiment to the point in the parameter space that produces high yields and appreciable concentrations for the Sn$V^-$ at low temperature.

\section{CONCLUSION\label{sec:conc}}
In conclusion, we have proposed a class of pathways, with the prototypical pathway labeled as $p_1$, that leads both to a high yield and to an appreciable defect concentration for Sn$V^-$. In particular, our proposed pathways suggest fluences of Sn that should maximize Sn$V^-$ yield in diamond growth experiments. A low chemical potential $\mu_\text{Sn}$ would most closely correspond to a low Sn fluence, while a high chemical potential $\mu_\text{Sn}$ would most closely correspond to a high Sn fluence. We believe our results would be very useful to experimentalists looking to incorporate solid state single-photon emitters into wide-bandgap hosts for quantum information and optoelectronic applications.

\section*{ ACKNOWLEDGMENTS:}
R.K.D. gratefully acknowledges financial support that made this work possible from the National Academies of Science, Engineering,
and Medicine Ford Foundation Postdoctoral Fellowship program, and from Syracuse University College of Engineering and Computer Science funds. We also acknowledge support by the STC Center for Integrated Quantum Materials, NSF Grant No. DMR-1231319.  

\section*{REFERENCES:}
\bibliography{refs_SnV-Carriers}
\end{document}